\documentclass[%
 reprint,
 amsmath,amssymb,
 aps,
 prx,
floatfix,
]{revtex4-2}

\usepackage[mode=buildnew]{standalone}
\usepackage{graphicx}
\usepackage{dcolumn}
\usepackage{bm}
\usepackage{braket}
\usepackage{xcolor}
\usepackage[utf8]{inputenc}
\usepackage[T1]{fontenc}
\usepackage[english]{babel}
\usepackage{letltxmacro}
\usepackage{mathtools}
\usepackage{ifthen}
\usepackage{ragged2e}
\usepackage{placeins}
\usepackage{enumerate}
\usepackage{siunitx}
\usepackage{aligned-overset}

\usepackage{tikz-cd}
\usepackage{tikz}
\usetikzlibrary{automata,arrows,positioning,calc,fit,arrows.meta,shapes}

\LetLtxMacro{\ORIGselectlanguage}{\selectlanguage}
\makeatletter
\DeclareRobustCommand{\selectlanguage}[1]{%
  \@ifundefined{alias@\string#1}
    {\ORIGselectlanguage{#1}}
    {\begingroup\edef\x{\endgroup
       \noexpand\ORIGselectlanguage{\@nameuse{alias@#1}}}\x}%
}
\newcommand{\definelanguagealias}[2]{%
  \@namedef{alias@#1}{#2}%
}

\definelanguagealias{en}{english}

\newcommand{\Path}[1]{\mathcal{P} \ifthenelse{\equal{#1}{}}{}{[#1]}}                    
\newcommand{\Pathr}[1]{\widetilde{\mathcal{P}}\ifthenelse{\equal{#1}{}}{}{[#1]}}        
\newcommand{\mean}[1]{\left\langle #1 \right\rangle}                                    
\newcommand{\obs}{\mathcal{O}}                                                          
\newcommand{\mev}[1]{\mathcal{#1}}                                                      
\newcommand{\ab}{\tau}                                                                  
\newcommand{\rev}[1]{\widetilde{#1}}                                                    
\newcommand{\ie}{\textit{i.e.}}                                                         

\newcommand{\re}[1]{\textcolor{black}{#1}}                                              

\newcommand{\traj}{\Gamma}                                                              
\newcommand{\Pathb}[1]{\mathcal{P}^\text{cf} \ifthenelse{\equal{#1}{}}{}{[#1]}}          
\newcommand{\todo}[1]{{\color{red}\textbf{TODO}\ifthenelse{\equal{#1}{}}{}{\textbf{:} #1}}}      
\newcommand{\period}{T}                                                                 
\newcommand{\kb}{k_\text{B}}                                                            

\newcommand{\SIref}[1]{{\color{blue}SI.#1}}                                              
\renewcommand{\eqref}[1]{[\ref{#1}]}
\renewcommand{\thesection}{\arabic{section}}

\makeatletter
\renewcommand{\p@subsection}{\thesection}
\makeatother

\begin{document}
\preprint{APS/123-QED}

\title{General theory for localizing the where and when of entropy production meets single-molecule experiments}

\author{Julius Degünther}
\thanks{J.D. and J.v.d.M. contributed equally to this work.}
\author{Jann van der Meer}
\thanks{J.D. and J.v.d.M. contributed equally to this work.}
\author{Udo Seifert}
\affiliation{
 II. Institut für Theoretische Physik, Universität Stuttgart, 70550 Stuttgart, Germany
}

\date{\today}

\begin{abstract}
The laws of thermodynamics apply to biophysical systems on the nanoscale as described by the framework of stochastic thermodynamics. This theory provides universal, exact relations for quantities like work, which have been verified in experiments where \re{ a fully resolved description allows direct access to such quantities}. Complementary studies consider partially hidden, coarse-grained descriptions, in which the mean entropy production typically is \re{not directly accessible but can be} bounded in terms of observable quantities. Going beyond the mean, we introduce a fluctuating entropy production that applies to individual trajectories in a coarse-grained description under time-dependent driving. Thus, this concept is applicable to the broad and experimentally significant class of driven systems in which not all relevant states can be resolved. We provide a paradigmatic example by studying an experimentally verified protein unfolding process. As a consequence, the entire distribution of the coarse-grained entropy production rather than merely its mean retains spatial and temporal information about the microscopic process. In particular, we obtain a bound on the distribution of the physical entropy production of individual unfolding events.
\end{abstract}

\maketitle

\section{Introduction}
Since its inception, stochastic thermodynamics gradually matured from a deductive framework to an inductive instrument. Its original purpose to identify and relate thermodynamic quantities along fluctuating trajectories of driven systems \cite{seki10, jarz11, seif12} has convincingly been demonstrated with experiments on colloidal systems and electronic circuits, as reviewed in Refs. \cite{cili17, peko15}. In biophysics, the first experiments along this line mainly investigated universal relations like, \emph{e.g.}, the Jarzynski or Crooks relation in small model systems that were fully accessible \cite{liph02, coll05, rito06, haya10}. At that time, the focus was on fundamental aspects like the relation of non-equilibrium measurements of work to equilibrium quantities like free energy. In contrast, the more recent inductive direction of studies explores the idea that thermodynamic reasoning can reveal hidden properties of a system on the basis of suitable observations. Molecular motors operating at high efficiency \cite{toya10, li20} and phenomena like currents of a given precision \cite{bara15, ging16, horo20}, coherent oscillations \cite{cao15, ober22, ohga23}, transport of distributions \cite{aure11, shir18, ito20, vu23}, particular patterns in distributions of transition paths or waiting times \cite{bere19, glad19, skin21a} and collective effects in active matter \cite{fodo22} require non-equilibrium, thus stimulating the idea that departure from equilibrium can be quantified through such observations.

The arguably most convincing quantifier for non-equilibrium is the physical entropy production, which the framework of stochastic thermodynamics describes as a fluctuating quantity that depends on the microscopic trajectory of all relevant degrees of freedom. Since it is impossible to access such a quantity directly beyond particularly simple, fully accessible models, a broad variety of tools has been developed to recover or at least bound the entropy production based on incomplete data. These theoretical methods include bounds based on precision, see, \emph{e.g.}, \cite{hwan18, li19, vu20, koyu20, dech21}, correlation times \cite{dech23}, counting events \cite{piet23} or extreme value statistics \cite{neri23}. A major class of results relies on interpreting the mean entropy production as a Kullback-Leibler divergence \cite{cove06, kawa07, rold10} to obtain fluctuation relations and entropy estimators based on incomplete data \cite{shir14, bisk17, mart19}. This technique is conceptually related to recently introduced frameworks for coarse graining  that include waiting-time distributions \cite{hart21a, vdm22, haru22, vdm23, degu23}. Contrasted with state lumping, which is the usual paradigm for coarse graining \cite{espo12, seif19}, these recent developments show novel, surprising phenomena \cite{hart23, blom23} and comprise a promising new direction for single-molecule studies \cite{gode23}. 

While most of the tools described above apply to the steady state or at least to a time-independent set-up, the description of experimental scenarios that include external driving inevitably requires an explicitly time-dependent framework. Beyond merely estimating entropy production, the reconstruction of entire energy landscapes of single molecules \cite{humm01, gupt11} relies on measurements both in and out of equilibrium, the latter performed by exerting a force via optical tweezers, as reviewed in Refs. \cite{wood14, camu16, poli18, bust21, gies21}.

In this work, we identify entropy production for coarse-grained time-dependent systems, \ie, for set-ups that include external driving and hidden degrees of freedom. Our approach describes the coarse-grained entropy production as a fluctuating quantity much like its microscopic counterpart. However, the coarse-grained entropy production can be determined from the available data alone. Thus, different microscopic realizations of the same observations yield the same coarse-grained entropy production, although the actual microscopic entropy production might differ in each individual realization. In this sense, the coarse-grained entropy production is, unlike its microscopic counterpart, ``operationally accessible''. \re{Even if the underlying dynamics is inaccessible, identifying this quantity is possible by verifying that some of the observations qualify as ``Markovian events'', which is a structural property that can be checked based on the coarse-grained data alone, as discussed below.} 

Moreover, the introduced coarse-grained entropy production is embedded into the broader context of stochastic thermodynamics and, in particular, yields a bound on the actual physical entropy production. On a technical level, these properties are established by generalizing the framework developed in Refs. \cite{vdm23, degu23} for stationary systems to time-dependent driving.

We will illustrate the main features of our approach for \re{a simple model} of a realistic protein unfolding experiment \cite{stig11} in which the externally controllable pulling force is varied. Since the entropy production we consider is a random variable, we are able to study its distribution in terms of different observables like, \textit{e.g.}, the starting point or duration of an unfolding event of a small protein. As a further consequence, we obtain bounds on the distribution of the microscopic entropy production, which is not directly accessible. \re{Since such bounds on distributions are statistically more demanding than estimates of mean quantities, in particular in a time-dependent setting, we also demonstrate how our method allows different degrees of resolution for the coarse-grained entropy production. In particular and going beyond the mean value, we obtain the distribution of this quantity in terms of one and two parameters.}


\section{Markovian events and their detection}

\subsection{\re{Microscopic and coarse-grained set-up}}
\label{sec:conceptual}

\begin{figure}[t]
\begin{center}
    \includegraphics{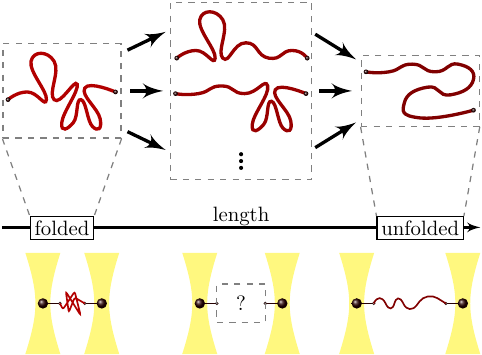}
    \caption{Sketch of a folding experiment. The molecule undergoes conformational changes between a folded, an unfolded and multiple intermediate states. The rate for these changes depends on the pulling force applied via optical tweezers acting on beads that are connected to each end of the molecule. The primary observable is the length of the molecule, which might prevent the unique identification of states with similar length.}
    \label{fig:fig0}
\end{center}
\end{figure}

We begin by illustrating the type of problem to which our results apply. The main goal is to reconstruct thermodynamic properties, in particular regarding the irreversibility of folding processes of, \textit{e.g.}, proteins or hairpin molecules, from incomplete data. Such data can be obtained from experiments where external manipulations affect which conformation is the most likely. In these setups, optical tweezers are used to exert a force on the molecule via beads attached to both of its ends, as sketched in the lower part of Figure~\ref{fig:fig0}. The primary measurement is the total length of the molecule at each point in time, which serves as a projected one-dimensional coordinate for its conformational state. 

Typically, such data provides an incomplete picture. In the case of calmodulin \cite{stig11}, measuring the length of the molecule is sufficient to identify a discrete folding network of six conformational states that captures the essential features. However, for a different protein we might have the less favorable case that, \textit{e.g.}, there are too many internal states or that states become indistinguishable because the length of the protein is within a too small range for these states. Thus, we consider a coarse-grained scenario in which only certain internal states can be resolved. In particular, we consider the case in which only the fully folded and the fully unfolded state can be uniquely identified through the shortest and longest measured length, respectively, while intermediate states cannot be distinguished, as shown in the upper part of Figure~\ref{fig:fig0}.

If the underlying, typically not fully accessible system can be described by a Markov network, the framework of stochastic thermodynamics allows a meaningful identification of entropy production in terms of the microscopic trajectory. In a coarse-grained system, such a microscopic definition of entropy production is usually not operationally accessible. For example, entropy production is well-defined for a discrete Markovian dynamics, but, as a Markovian dynamics does not remain Markovian under coarse graining, the usual definition of entropy production is not immediately applicable. One approach to recover a meaningful notion of entropy production is to identify Markovian events, a concept introduced for a time-independent set-up in Refs. \cite{vdm23, degu23}. Put loosely, we try to structure the coarse-grained dynamics so that memory effects are limited to the transition region from one Markovian event to the following one, without affecting subsequent transitions across multiple Markovian events. 

The purpose of this paper is to generalize this concept to time-dependently driven systems, where some states in a Markov network can be observed while others cannot, as illustrated in Figure~\ref{fig:fig0} and Figure~\ref{fig:cal_protocol}a) and b). The remainder of this section \re{describes the framework of Markovian events in detail and discusses criteria to identify such events in practice}. Thereafter, we discuss and illustrate applications \re{of this concept} to protein folding experiments in the Sections~\ref{sec:calmodulin} to \ref{sec:boundcdf}. Proofs and \re{details on the numerical results} are found in the SI. 

\subsection{\re{Markovian events and snippets}}
\label{sec:MEV_snip}

\re{Our results utilize the concepts of Markovian events and snippets introduced in Refs. \cite{vdm23, degu23} for the simpler case of time-independent driving. Here, we assume that the microscopic time-evolution of the system can be described by a trajectory $\gamma(t)$, with $0 \leq t \leq T$. If the underlying dynamics is Markovian, knowing $\gamma(\tau)$, \ie, the state of the system at time $\tau$, includes all relevant information for the future time-evolution of $\gamma_+$, the part of the trajectory for $t \geq \tau$. We can formulate this Markov property as a conditional independence condition
\begin{equation} \label{eq:mp}
    \Path{\gamma_+|\gamma(\tau), \gamma_-} = \Path{\gamma_+|\gamma(\tau)}
\end{equation}
between $\gamma_+$ and the past time-evolution $\gamma_-$ given the current state $\gamma(\tau)$. \re{In this equation, the symbol $\Path{}$ denotes the path weight, which serves as a probability density for trajectories.} If the dynamics is not Markovian due to memory effects of some sort or if it is not fully accessible due to practical limitations, Equation~\eqref{eq:mp} may not be immediately operationally useful. However, it is conceivable that a property similar to Equation~\eqref{eq:mp} is tied to some particular observation $I$ at time $\tau$ and therefore holds at some but not all points in time. Thus, we define a Markovian event as a tuple
\begin{equation}
    \mev{I}\equiv(I,\ab)
\end{equation} 
that satisfies
\begin{equation} \label{eq:mev}
    \Path{\gamma_+|\mev{I}, \gamma_-} = \Path{\gamma_+|\mev{I}}
.\end{equation}
Examples include the observation of a Markov state or the detection of transitions between such states in a system where the remaining parts may not be observable.}

\re{The time-evolution of $\gamma$ may contain dynamical memory effects in the transition region between two Markovian events, so that it makes sense to treat small sections of a long trajectory between two consecutive Markovian events as a dynamical unit. We refer to such trajectory sections as snippets. We can symbolically denote a snippet between the two Markovian events $\mev{I}$ and $\mev{J}$ by
\begin{align} \label{eq:snip}
    \traj:\quad\quad\mev{I}&\xrightarrow[]{t}\mev{J}\\
    (I, \tau)&\xrightarrow[]{}(J, \tau+t)
,\end{align}
where $t$ is the duration of the snippet. In fact, it is possible to include additional observations $\obs$ that are not Markovian events in a snippet in the form
\begin{align} \label{eq:snip_b}
    \traj:\quad\quad\mev{I}&\xrightarrow[]{t,\obs}\mev{J}
,\end{align}
which was discussed in Refs. \cite{vdm23, degu23} for the time-independent case. For simplicity, we avoid such additional notation in the following. Note that a snippet is denoted with a capital letter $\traj$ to indicate that it is defined on the observable, coarse-grained level, whereas a microscopic trajectory is denoted with a small letter $\gamma$.}

\subsection{\re{Time-dependent driving and time reversal}}
\label{sec:td_driving}

\re{For a system driven by a time-dependent protocol $\lambda(t)$ from $t=0$ to $t=T$, identifying the microscopic entropy production requires knowledge of both the original system and of the system driven by the corresponding time-reversed protocol $\rev{\lambda}(t)=\lambda(T-t)$  \cite{seif12}. To discern these physically distinct protocols, we denote probabilities and path weights by $P$ and $\Path{}$ for the original driving and $\rev{P}$ and $\Pathr{}$ for the time-reversed driving, respectively.}

\re{If a Markovian event $\mev{I}$ is associated with an observation $I$ that transforms under time reversal according to $I \mapsto \tilde{I}$ with $\tilde{I} \neq I$, which applies, \textit{e.g.}, to a momentum variable or to a transition, we write 
\begin{equation}
    \mev{I}=(I,\tau) \mapsto \rev{\mev{I}} = (\rev{I},\tau)
\end{equation}
to indicate that external control parameters governed by the protocol $\lambda$ take the same value at the time of the event $\mev{I}$ and its time reverse $\rev{\mev{I}}$, since $\lambda(\ab) = \tilde{\lambda}(T-\ab)$.}

\subsection{\re{How to identify Markovian events}}

\re{The identification of Markovian events requires sufficient structure in the coarse-grained data. From a practical perspective, verifying the defining property \eqref{eq:mev} directly does not seem promising since one would need path weights of trajectories on the microscopic level. However, Markovian events can be identified by other means, as will be detailed in the following.}

\re{In some cases, particular events on the coarse-grained trajectory can be classified as Markovian due to reasoning that refers to the underlying physics. For example, in fluorescence resonance energy transfer (FRET) experiments it is possible to observe particular transitions which are accompanied by the emission of a detectable photon \cite{gode23} giving rise to a clean Markovian event. Another possibility is to justify Markovian events based on the time scales relevant for the process. While dynamics on a Markov network implicitly assumes instantaneous transitions between its states, the requirements to identify a Markovian event are more flexible and can take into account that transitions are not infinitely fast \cite{sati20, hart21a}. For example, we can treat a particular coarse-grained state as Markovian if transitions within this state are much faster than transitions into or out of this state so that the state equilibrates locally. We emphasize that this property can be checked for each state individually, so that Markov states can be found even if a time-scale separation is not granted for the entire microscopic dynamics. Consequently, this framework is more flexible than, e.g., assuming an underlying, not fully observable Markov network which would include imposing the Markov assumption on its hidden parts.}

\re{Apart from such constructive arguments that rely on knowledge about the mechanisms involved, it is also possible to formulate criteria that the Markov property \eqref{eq:mev} is violated. Since in the time-independent case a state that satisfies the Markov assumption must have an exponential waiting-time distribution, a stationary measurement of the lifetimes of a particular state provides evidence in favor of or against treating this state as a Markov state. The method used to obtain the folding network of calmodulin \cite{stig11} can be sorted into this category. By measuring the protein length, it is possible to distinguish four different predominant lengths. Three of these four tentative states (including the fully folded and fully unfolded state) feature an exponential distribution of their lifetime for constant external force and can therefore be justified as Markov states. The remaining three states share a similar protein length and were distinguished through hidden-Markov analysis. We expect that such incomplete access to the lifetimes is indeed the generic case in more complex scenarios, so that some states still give rise to Markovian events even if the underlying network cannot be reconstructed.}

\re{A conceptually related, but more abstract criterion has been put forward in Ref. \cite{vdm23} for the time-independent case. If, for example, a non-Markov state is erroneously treated as a Markov state, cutting at every $n$-th occurrence of this state gives a bias in the resulting coarse-grained entropy production that depends on $n$. Generalizing this argument to the time-dependent case, we obtain an operational criterion to check whether some state/event satisfies the Markov property or not. On a related note, criteria that are able to rule out that the Markov assumption is valid for given trajectory data have been suggested, e.g., by analyzing the statistics of transition paths \cite{bere18} or the deviation from a reference Markov process \cite{lapo21}.}


\section{Results}
\label{sec:results}

\subsection{\re{Coarse-grained and microscopic entropy production}}

\re{Given a microscopic trajectory $\gamma$ initialized and finalized by the Markovian events $\mathcal{I}$ and $\mathcal{J}$, respectively, and the reverse trajectory $\tilde{\gamma}$, we identify the time-dependent microscopic entropy production of $\gamma$ as
\begin{equation}
    \Delta s[\gamma] = \kb\ln \frac{P(\mev{I})\Path{\gamma|\mev{I}}}{P(\mev{J})\Pathr{\rev{\gamma}|\rev{\mev{J}}}}
    \label{eq:micro_delta_s}
.\end{equation}
This identification becomes the usual expression for a time-dependent entropy production in Markovian dynamics if $\mathcal{I}, \mathcal{J}$ denote odd or even state variables \cite{seif12, spin12} or transitions \cite{degu23}. On the coarse-grained level, the entropy production of a snippet $\traj =\mev{I}\xrightarrow{}\mev{J}$ can be identified in a similar fashion as 
\begin{equation} \label{eq:Delta_S_TD} 
    \Delta S[\traj]=\kb\ln\frac{P(\mev{I})\Path{\traj|\mev{I}}}{P(\mev{J})\Pathr{\rev{\traj}|\rev{\mev{J}}}}
,\end{equation}
which generalizes the identification in \cite{degu23} to time-dependently driven processes. Typically, the snippet $\Gamma$ is a small section of a trajectory that is embedded in a longer one.}
\re{We emphasize that, in analogy to Equation~\eqref{eq:micro_delta_s}, the probability $P(\mev{J})$ in the denominator originates from the forward process. If the Markovian events $\mev{I}, \mev{J}$ themselves are discontinuous and contribute to entropy production singularly, the convention~\eqref{eq:Delta_S_TD} includes the initial contribution due to $\mev{I}$ but not the final one due to  $\mev{J}$. In this case, the defining property of the Markovian events, Equation~\eqref{eq:mev}, must hold immediately prior to events $\mev{I}$ and $\mev{J}$ to account for the singular contributions correctly. This peculiar case is relevant for models featuring a discrete dynamics, which includes the unfolding process discussed in Section~\SIref{1.2}.}

\subsection{\re{Estimation of entropy production on the trajectory level}}

As we prove in the SI, Equation~\eqref{eq:Delta_S_TD} is consistent with the usual identification of a time-dependent entropy production $\Delta s[\gamma]$ on the level of the microscopic trajectory $\gamma$ \cite{seif12}. For example, averaging this coarse-grained entropy production results in a meaningful lower bound on the mean physical entropy production $\mean{\Delta s}$,
\begin{equation} \label{eq:Delta_S_bound}
    0 \leq \mean{\Delta S} \leq \mean{\Delta s}
.\end{equation}
In particular, if the coarse-grained process is not in apparent equilibrium, \ie, if $\mean{\Delta S} > 0$, we can rule out a genuine equilibrium on the microscopic level. Thus, the entropy estimator is consistent from a thermodynamic perspective. The bound~\eqref{eq:Delta_S_bound} only makes use of the expectation value of the fluctuating quantity $\Delta S[\Gamma]$, whereas its full distribution contains more information. In fact, we can establish 
\begin{equation}
    \Delta S[\Gamma] \leq \mean{\Delta s|\Gamma}
    \label{eq:Delta_S_lower}
,\end{equation}
which, compared to the inequality~\eqref{eq:Delta_S_bound}, provides not a single bound, but a different one for each $\Gamma$. Since the microscopic entropy is averaged over all realizations that produce a particular coarse-grained trajectory section $\Gamma$, this inequality is formulated on a much finer scale. We stress that this result is independent of model specifications and, given an underlying Markovian dynamics, holds for any choice of rates and any topology under arbitrary time-dependent protocols.


\subsection{Coarse-grained driven unfolding process of calmodulin as paradigm}
\label{sec:calmodulin}

\begin{figure}[t!]
\begin{center}
    \includegraphics[scale=1]{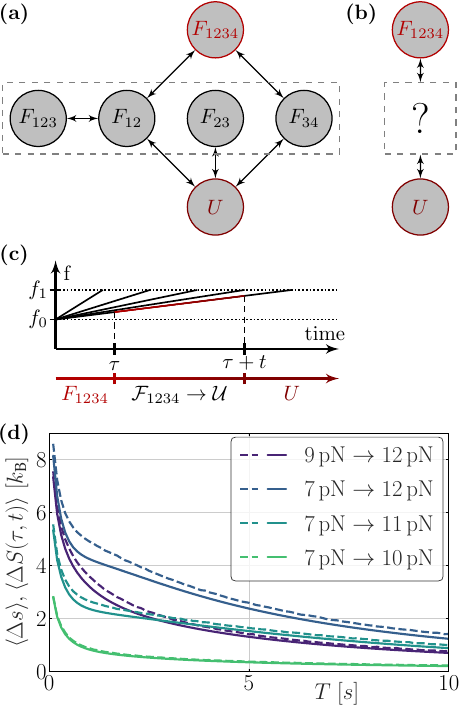}
    \caption{From a partially observed network to isolating entropy production. (a) Six-state Markov network of calmodulin. The labeling follows Ref.~\cite{stig11}. \re{The rates are given by Equation~\eqref{eq:cal_rate_param} with the concomitant parameters listed in Section~\SIref{4}.} We assume that only the fully folded and the fully unfolded state can be uniquely identified in accordance with the scenario sketched in Figure~\ref{fig:fig0}. (b) Coarse-grained description of the network shown in a). (c) Driving the system. The force ramp increases linearly from $f_0$ to $f_1$ in time $T$. An unfolding event leaves the folded state $F_{1234}$ at time $\tau$ and enters the unfolded state $U$ after duration $t$. (d) Average entropy production of an unfolding event as a function of the protocol duration $T$, plotted for different values of $f_0$ and $f_1$. The dashed line shows the average microscopic entropy $\Delta s_\text{unf}$ for an unfolding event, which can be calculated with access to the microscopic dynamics. The solid line depicts the coarse-grained entropy production $\Delta S_\text{unf}$ as an estimator that provides a lower bound, which can be determined from the coarse-grained data. In the figure, we omit the subscript ``unf''.}
    \label{fig:cal_protocol}
\end{center}
\end{figure}

We now illustrate how the theoretical results introduced above can be implemented in practice. As a concrete application, we consider simulations based on the protein folding experiment performed by Stigler \textit{et al.} \cite{stig11}. The main result of this experiment is the identification of a folding network for calmodulin that comprises six internal Markov states as shown in Figure~\ref{fig:cal_protocol}a). \re{All transition rates between the six conformational states depend on the force externally applied via optical tweezers.} For forces in the range of \qty{6}{\pico\newton} to \qty{13}{\pico\newton}, the rate from state $i$ to state $j$ is approximately given by
\begin{equation} \label{eq:cal_rate_param}
    k_{ij}=k_{0, ij}e^{\kappa_{ij}f}
\end{equation}
with the experimentally determined parameters $k_{0, ij}$ and $\kappa_{ij}$~\cite{stig11, koyu20} as listed in Section~\SIref{3} in the SI. The network shown in Figure~\ref{fig:cal_protocol}a) and the rates~\eqref{eq:cal_rate_param} \re{provide the parameters for the model based on the experiment reported in \cite{stig11}.}

We assume that only the folded state $F_{1234}$ and the unfolded state $U$ can be resolved, while the remaining partially or misfolded states cannot be distinguished, which leads to the effective network shown in Figure~\ref{fig:cal_protocol}b). Thus, we have a concrete instance for the situation detailed in Section~\ref{sec:conceptual}. To simulate an experimental scenario in which external driving is used to induce the unfolding process, the system is initialized in the equilibrium state corresponding to $f(0)=f_0$ and subsequently subjected to the linearly increasing force
\begin{equation} \label{eq:protocol_calmodulin}
    f(t)=f_0 + (f_1-f_0)t/T
,\end{equation}
with the total length of the driving $T$ and the final force $f(T)=f_1$. 

\subsection{Where? - Estimating entropy production of unfolding processes with different protocols}
\label{sec:numerical1}

We aim at a bound on the entropy production of unfolding events that occur during the protocol~\eqref{eq:protocol_calmodulin}. Using the notation introduced in Figure~\ref{fig:cal_protocol}a), these realizations start with the Markovian event $\mev{F}_{1234}$, which denotes the exit of state $F_{1234}$ at time $\ab$, and end with $\mev{U}$, the entry into state $U$ at $\tau+t$, without sojourn in $F_{1234}$ or $U$. These times have to fulfill $0\leq\ab\leq T$ and $\ab\leq\ab+t\leq T$. Based on Equation~\eqref{eq:Delta_S_TD}, we identify the coarse-grained entropy production of an unfolding process
\begin{equation} \label{eq:S_unfolding}
    \Delta S_\text{unf}(\ab, t) \equiv \Delta S[(\text{exit } F_{1234}, \tau) \to(\text{enter } U, \tau+t)],
\end{equation}
as further detailed in Section~\SIref{2A} in the SI. Generally, $\Delta S_\text{unf}(\ab, t)$ is a fluctuating quantity that changes with each realization characterized by the two time variables $\ab$ and $t$.

\begin{figure*}[t!]
\begin{center}
    \includegraphics[scale=1]{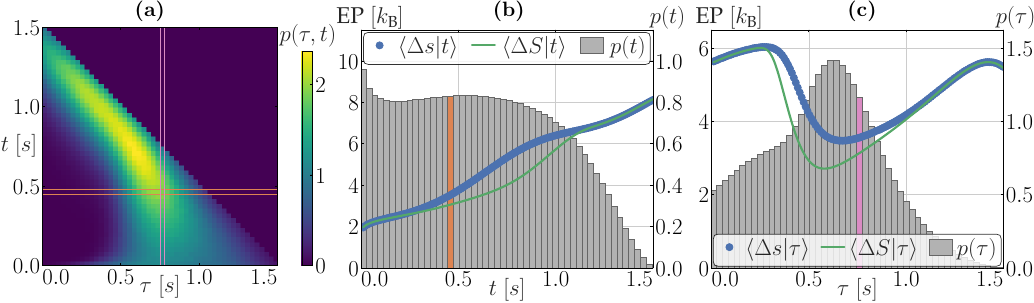}
    \caption{\re{Localizing entropy production in time. (a) Probability distribution $p_\text{unf}(\tau, t)$ for unfolding events starting at time $\ab$ with duration $t$. 
    Unfolding events can be classified in terms of, \textit{e.g.}, one of these two variables, the duration $t$ or the starting time $\tau$ as indicated by the orange and pink bars, respectively. (b)~Averaged entropy production for unfolding events of different duration $t$. Averaging over one row in a) leads to one data point/bar in b). The orange bar corresponds to the region between the two horizontal orange lines in a) and indicates how the unfolding events are grouped. For any possible duration $t$, averaging the coarse-grained entropy production over the respective realizations yields $\mean{\Delta S_\text{unf}|t}$, which underestimates the corresponding average $\mean{\Delta s_\text{unf}|t}$ of the microscopic entropy production. Access to $\mean{\Delta S_\text{unf}|t}$ as well as the corresponding probability $p(t)$ requires only coarse-grained data, whereas the reference value $\mean{\Delta s_\text{unf}|t}$ is computed from microscopic model details. (c)~Averaged entropy production $\mean{\Delta S_\text{unf}|\tau}$ for unfolding events starting at different times $\tau$ and probability distribution $p(\tau)$ for this starting time $\tau$. Again, a lower bound $\mean{\Delta S_\text{unf}|\ab}\leq\mean{\Delta s_\text{unf}|\tau}$ is obtained. The pink bar corresponds to the region between the two vertical pink lines from a) and indicates how the unfolding events are grouped. Averaging over one column in a) leads to one data point/bar in c). The parameters of the system used in this figure are $T=\qty{1.5}{\sec}$, $f_0=\qty{7}{\pico\newton}$ and $f_1=\qty{12}{\pico\newton}$. We omit the subscript ``unf'' and use EP as abbreviation for entropy production.}}
    \label{fig:cal_time}
\end{center}
\end{figure*}	

A fluctuating coarse-grained entropy production of the form of Eq.~\eqref{eq:S_unfolding} or, more abstractly, of Eq.~\eqref{eq:Delta_S_TD} allows us to attribute entropy production to different paths. An identification on the level of small, local trajectories $\mev{I} \to \mev{J}$ between two Markovian events $\mev{I}$, $\mev{J}$ also amounts to knowing where the entropy production actually happens. Returning to our set-up, we are able to distinguish unfolding $F_{1234}\to U$ from folding $U\to F_{1234}$ and from unproductive loops $F_{1234}\to F_{1234}$ and $U\to U$. In a time-dependent scenario, even such loops contribute to entropy production, as further detailed in Section~\SIref{3}.



We analyze the influence of the protocol parameters $f_0$, $f_1$ and $\period$ on the entropy production of an unfolding event. More specifically, we consider the mean coarse-grained entropy production of an unfolding event with a given protocol $\mean{\Delta S_\text{unf}(\ab, t)}$, which results from a weighted average over all possible realizations as further detailed in Section~\SIref{2B} in the SI and compare it to its microscopic counterpart $\mean{\Delta s_\text{unf}}$. Note that these averages depend on the specific choice of the protocol, \ie, the parameters $f_0$, $f_1$ and $\period$.

Figure~\ref{fig:cal_protocol}d) shows the results of the entropy estimation for a range of different total lengths of the driving $T$ and for several combinations of the force parameters $f_0$ and $f_1$. Overall, we find that the coarse-grained entropy production $\mean{\Delta S_\text{unf}(\ab, t)}$ recovers a significant part of the corresponding microscopic entropy production and correctly reproduces the main features of the curve. In particular, we find that fast protocols lead to unfolding events with larger entropy production.

We stress that such detailed inference is not possible by merely knowing the mean entropy production, which does not allow one to resolve between, \textit{e.g.}, folding and unfolding or, more generally, different pathways.
\re{Moreover, apart from identifying appropriate Markovian events like the folded and unfolded state applying our theoretical framework does not require knowledge of the network topology or transition rates.} We use these exclusively \re{to construct a model based on realistic experimental results} and for obtaining the full, microscopic entropy production as a benchmark. All quantities that are necessary to apply our methods, as illustrated here and in the following sections, are experimentally accessible by observing the folded and unfolded state under the original and the time-reversed protocol without any further input. In particular, these methods will apply to data from any protein in which only certain states are observable even if the entire folding network is unknown \re{and even if the driving happens within this unknown part only. For the time-independent case, Ref. \cite{degu23} demonstrates that even driven cycles that are completely located in the hidden part of the network can be inferred.} 

\subsection{When? - Localizing entropy production in time}
\label{sec:numerical2}

So far, we used the ability of our framework to localize the irreversibility in state space, \textit{i.e.}, we isolate the entropy production that originates from unfolding processes from that of other events like unproductive loops that also contribute to the total entropy production of the system. To further dissect the irreversibility of these unfolding events, we now exploit the ability to localize entropy production in time.

In this spirit, we first raise the question whether the entropy production for a given overall protocol duration $\period$ primarily stems from fast unfolding events or slow ones, as illustrated in Figure~\ref{fig:cal_time}a). Thus, we consider the average entropy production of the calmodulin unfolding conditioned on the duration of the unfolding $\mean{\Delta S_\text{unf}|t}$ and compare it to its microscopic counterpart $\mean{\Delta s_\text{unf}|t}$. To fully assess the contributions of fast and slow events, we additionally consider the probability to observe an unfolding event of duration $t$
\begin{equation}
    p_\text{unf}^\text{dur}(t)\equiv\sum_{\traj|\text{duration }t}\Path{\traj|\text{unfolding}},
\end{equation}
where the sum runs over all corresponding coarse-grained unfolding trajectories.

For the example shown in Figure~\ref{fig:cal_time}b), we find good agreement between the estimate and the real entropy production, especially for the limiting cases of short and long unfolding events. In particular, we correctly identify that slow unfolding on average produces more entropy but occurs less often than fast unfolding. \re{Slow unfolding events tend to contain more jumps due to transitions between $F_{12}$ and $F_{123}$. Since the system is time-dependently driven, such longer trajectories on average lead to a larger entropy production. The rate of entropy production of an unfolding event ${\mean{\Delta S_\text{unf}|t}/t}$, however, generally decreases with increasing duration $t$.}

Second, we compare events that start earlier or later during the force ramp protocol, as illustrated in Figure~\ref{fig:cal_time}b). Analogously, this irreversibility is quantified by the average entropy production of the unfolding conditioned on the initial time of the event $\mean{\Delta S_\text{unf}|\tau}$, which we compare to its microscopic counterpart $\mean{\Delta s_\text{unf}|\tau}$. The corresponding probability distribution is given by
\begin{equation}
    p_\text{unf}^\text{ini}(\tau)\equiv\sum_{\traj|\text{start at }\ab}\Path{\traj|\text{unfolding}},
\end{equation}
where the sum runs over all coarse-grained unfolding trajectories that begin at time $\ab$.

Figure~\ref{fig:cal_time}c) reveals that the average entropy production is lowest for events that start halfway through the protocol, which are also the most frequent ones. Late unfolding events are particularly rare since the condition $\tau+t\leq\period$ then necessarily requires them to be fast. Again, the estimator $\mean{\Delta S_\text{unf}|\tau}$ reproduces the main features of the real entropy production $\mean{\Delta s_\text{unf}|\tau}$.

\re{From a practical perspective, in future experiments it may be possible that the data does not permit results as detailed as the one shown in Figure~\ref{fig:cal_time} with sufficient statistical significance. In this case, it is possible to resort to a coarser resolution of the histogram or, in the extreme case, even to a binary classification of, for example, early and late processes depending on whether the starting time $\tau$ surpasses a given threshold $\tau_0$ or not. Such binned results are still more expressive than the mere mean value.}

\begin{figure*}[t]
\begin{center}
    \includegraphics[scale=1]{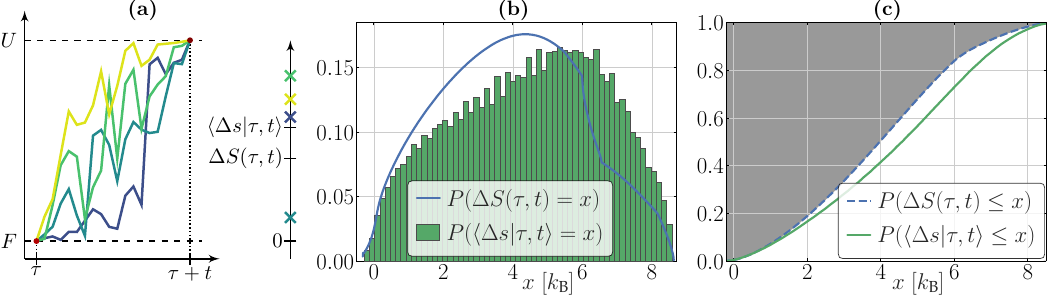}
    \caption{Entropy production as a random variable and a bound on the corresponding distributions. (a) \re{Sketch highlighting the relationship between entropy production $\Delta s$ on the microscopic level and $\Delta S$ on the coarse-grained level. The left diagram sketches different realizations of the dynamics of a microscopic variable such as the length of a protein as a function of time. Given a coarse graining in which the region between $F$ and $U$ cannot be resolved, the depicted microscopic trajectories give rise to the same coarse-grained unfolding trajectory that is characterized by its initial time $\tau$ at which the observable state $F$ is exited and its duration $t$ after which the observable state $U$ is entered. In accordance with the bound~\eqref{eq:Delta_S_lower}, the corresponding coarse-grained entropy production ${\Delta S(\tau,t)}$ underestimates the average value of the microscopic entropy production ${\mean{\Delta s|\tau,t}}$. } (b) Probability distribution functions of $\Delta S(\tau,t)$ and $\mean{\Delta s|\tau,t}$, with the latter as a histogram. (c) Cumulative distribution functions (CDFs) of $\Delta S(\tau,t)$ and $\mean{\Delta s|\tau,t}$. The relationship~\eqref{eq:Delta_S_lower} between $\Delta S(\tau,t)$ and $\mean{\Delta s|\tau,t}$ implies the bound~\eqref{eq:boundcdf} on the CDF of $\mean{\Delta s|\tau,t}$ in terms of the CDF of the coarse-grained entropy production. The function $P(S(\tau,t) \leq x)$ remains above $P(\mean{\Delta s|\tau,t} \leq x)$, so that the former, operationally accessible  distribution marks a no-go area for the latter. The parameters for the system used in b) and c) are $T=\qty{1.5}{\sec}$, $f_0=\qty{7}{\pico\newton}$ and $f_1=\qty{12}{\pico\newton}$. We omit the subscript ``unf''.}
    \label{fig:cal_distribution}
\end{center}
\end{figure*}


\subsection{Bound on the cumulative distribution}
\label{sec:boundcdf}

In this section, we derive and illustrate a universal bound on the distribution of $\Delta s$, which is valid for any model that qualifies for the framework described in \ref{sec:conceptual}. Equation~\eqref{eq:Delta_S_lower} implies that $\theta(x - \mean{\Delta s|\Gamma}) \leq \theta(x - \Delta S[\Gamma])$ holds true for all coarse-grained trajectories $\Gamma$, where the Heaviside function is defined by $\theta(y) = 1$ for $y\geq0$ and $\theta(y) = 0$ for $y<0$, respectively. For any fixed value of $x$, a summation over all $\Gamma$ yields the bound
\begin{equation} \label{eq:boundcdf}
    P \left( \mean{\Delta s|\Gamma} \leq x \right) \leq P \left( \Delta S[\Gamma] \leq x\right)
\end{equation}
between the two corresponding cumulative distribution functions. Thus, the distribution of the measurable, coarse-grained quantity $\Delta S[\Gamma]$ provides a bound on the cumulative distribution of the inaccessible microscopic entropy production. 

This result applied to the unfolding of calmodulin is shown in Figure~\ref{fig:cal_distribution}. We provide a graphical representation of the inequality~\eqref{eq:boundcdf} as an upper bound on the region that is accessible to the function $P \left( \mean{\Delta s_\text{unf}|\ab, t} \leq x \right)$. While the cumulative distributions shown in Figure~\ref{fig:cal_distribution}c) are linked by the relation~\eqref{eq:boundcdf}, the probability distributions $P(\Delta S_\text{unf}(\tau, t)=x)$ and $P(\mean{\Delta s_\text{unf}|\ab, t}=x)$ shown in Figure~\ref{fig:cal_distribution}b) obey no quantitative ordering.


\section{Discussion}
\label{sec:discussion}

In recent years, stochastic thermodynamics has developed a variety of tools and methods that rest on its comprehensive theoretical framework. From an experimental point of view, universal results like the Jarzynski equality~\cite{jarz11} that apply to time-dependent driving have been exploited successfully, whereas such an application is still pending for many of the more recent theoretical results in the field. One reason is that results are often formulated for restrictive model classes like, \textit{e.g.}, for a Markovian dynamics in the steady state. Another reason, which applies to inductive results for ``thermodynamic inference'' in particular, is that a genuine test of the methods is only possible if the quantity that is to be inferred, \textit{e.g.}, the microscopic entropy production, is accessible by other means, which in turn renders the coarse-graining unneccessary in the first place. 

For the particular system studied in this work all relevant parameters were determined experimentally. Consequently, access to both the coarse-grained and the underlying level allows a quantitative comparison of our results and an assessment of their quality. Nevertheless, the assumption that only the fully folded and unfolded state are visible has been motivated by the fact that such coarse graining is often inevitable in experiments. Since arbitrary time-dependent protocols are allowed, we expect that the generality of our approach invites experimental applications of a fluctuating coarse-grained entropy production.

\re{We emphasize that the crucial step in applying our results is the identification of appropriate Markovian states/events. Thus, it will also be of interest to find additional ways to test this Markov property in practice. Going beyond the criteria we discussed previously, we expect that known criteria that rule out that a process is Markovian like, e.g. the ones suggested in Refs. \cite{bere18, lapo21}, can be refined to obtain tests for non-Markovianity that can be applied to particular states/events.}

\re{For this specific model of calmodulin unfolding, we find that unfolding might not be the predominant type of process although the protocol is specifically designed to induce unfolding. Instead, the protein undergoes many unproductive loops if the driving is sufficiently slow. For the specific set of parameters used in Figure~\ref{fig:cal_time}, faster realizations of unfolding turn out to be more likely than slower ones and produce less entropy on average. Analogously, unfolding tends to start at an intermediate time of the protocol, which is also associated with a lower entropy production. The intermediate starting time corresponds to an externally applied force of roughly~\SI{9}{\pico\newton}. Such an analysis can also be repeated for similar protein folding experiments with other proteins and/or other forms of driving.}

It is important to note the conceptual difference between entropy production and work. In a force spectroscopy experiment, work can be determined from the applied external forces and the distance of the tweezers (see, \textit{e.g.}, Ref. \cite{humm01}), hence the coarse-grained data suffices to recover the work distribution. In contrast, this incomplete data does not yield the distribution of entropy production or even its value along an individual trajectory, which is defined microscopically. Our result implies a lower bound on this inaccessible distribution of entropy production, a substantially stronger result than an inequality between the respective expectation values, which is also implied. In special cases like a time-independent set-up \cite{degu23} or in case of full access to the system \cite{seif12}, the fluctuating entropy production for coarse-grained, driven systems identified in this manuscript reduces to established concepts. 

\re{It will also be worthwhile to explore whether it is possible to distinguish externally induced entropy production from internal dissipative processes like, e.g., the consumption of ATP. While such processes are not present in the discussed example of protein unfolding, such studies are certainly interesting for biomolecules like molecular motors that convert chemical energy. A theoretical approach to distinguish time-dependence due to external driving from steady internal activity is to investigate whether a decomposition of entropy production into housekeeping and excess contributions can be done on the coarse-grained level.}

From a theoretical standpoint, we may also wonder what happens if the requirements of this method are not met or are only met approximately. While the formalism includes discrete systems as well as overdamped and underdamped Langevin dynamics \cite{degu23}, systems with persistent memory effects do not fall into this category. For example, the situation discussed in Ref. \cite{blom23} considers a conceptually related effective theory for lumped states, \ie, a group of Markov states that when treated as a single entity are no longer Markovian. \re{Similarly, in Langevin dynamics of dimensions greater than one the probability of a point-like event approaches zero for an infinitely fine resolution \cite{dieb22}, whereas such events are not strictly Markovian for a coarse resolution. In this case, one would have to check Markovianity based on, e.g., local time-scale separation arguments or other criteria for the identification of Markovian events as discussed above. More generally,} it will be interesting to investigate such broader scenarios in which a fluctuating coarse-grained entropy production \re{retains its thermodynamic significance in an approximate sense} or, at least, \re{provides} an effective tool for inference even if a thermodynamic interpretation cannot be supported.

\textbf{Acknowledgments.} We thank Benjamin Ertel for fruitful discussions.

\bibliography{bib/references, bib/refs}

\begin{thebibliography}{62}%
\makeatletter
\providecommand \@ifxundefined [1]{%
 \@ifx{#1\undefined}
}%
\providecommand \@ifnum [1]{%
 \ifnum #1\expandafter \@firstoftwo
 \else \expandafter \@secondoftwo
 \fi
}%
\providecommand \@ifx [1]{%
 \ifx #1\expandafter \@firstoftwo
 \else \expandafter \@secondoftwo
 \fi
}%
\providecommand \natexlab [1]{#1}%
\providecommand \enquote  [1]{``#1''}%
\providecommand \bibnamefont  [1]{#1}%
\providecommand \bibfnamefont [1]{#1}%
\providecommand \citenamefont [1]{#1}%
\providecommand \href@noop [0]{\@secondoftwo}%
\providecommand \href [0]{\begingroup \@sanitize@url \@href}%
\providecommand \@href[1]{\@@startlink{#1}\@@href}%
\providecommand \@@href[1]{\endgroup#1\@@endlink}%
\providecommand \@sanitize@url [0]{\catcode `\\12\catcode `\$12\catcode
  `\&12\catcode `\#12\catcode `\^12\catcode `\_12\catcode `\%12\relax}%
\providecommand \@@startlink[1]{}%
\providecommand \@@endlink[0]{}%
\providecommand \url  [0]{\begingroup\@sanitize@url \@url }%
\providecommand \@url [1]{\endgroup\@href {#1}{\urlprefix }}%
\providecommand \urlprefix  [0]{URL }%
\providecommand \Eprint [0]{\href }%
\providecommand \doibase [0]{https://doi.org/}%
\providecommand \selectlanguage [0]{\@gobble}%
\providecommand \bibinfo  [0]{\@secondoftwo}%
\providecommand \bibfield  [0]{\@secondoftwo}%
\providecommand \translation [1]{[#1]}%
\providecommand \BibitemOpen [0]{}%
\providecommand \bibitemStop [0]{}%
\providecommand \bibitemNoStop [0]{.\EOS\space}%
\providecommand \EOS [0]{\spacefactor3000\relax}%
\providecommand \BibitemShut  [1]{\csname bibitem#1\endcsname}%
\let\auto@bib@innerbib\@empty
\bibitem [{\citenamefont {Sekimoto}(2010)}]{seki10}%
  \BibitemOpen
  \bibfield  {author} {\bibinfo {author} {\bibfnamefont {K.}~\bibnamefont
  {Sekimoto}},\ }\href {https://doi.org/10.1007/978-3-642-05411-2} {\emph
  {\bibinfo {title} {Stochastic Energetics}}}\ (\bibinfo  {publisher}
  {Springer},\ \bibinfo {address} {Berlin, Heidelberg},\ \bibinfo {year}
  {2010})\BibitemShut {NoStop}%
\bibitem [{\citenamefont {{J}arzynski}(2011)}]{jarz11}%
  \BibitemOpen
  \bibfield  {author} {\bibinfo {author} {\bibfnamefont {C.}~\bibnamefont
  {{J}arzynski}},\ }\bibfield  {title} {\bibinfo {title} {Equalities and
  inequalities: Irreversibility and the second law of thermodynamics at the
  nanoscale},\ }\href
  {https://doi.org/10.1146/annurev-conmatphys-062910-140506} {\bibfield
  {journal} {\bibinfo  {journal} {Ann. Rev. Cond. Mat. Phys.}\ }\textbf
  {\bibinfo {volume} {2}},\ \bibinfo {pages} {329} (\bibinfo {year}
  {2011})}\BibitemShut {NoStop}%
\bibitem [{\citenamefont {Seifert}(2012)}]{seif12}%
  \BibitemOpen
  \bibfield  {author} {\bibinfo {author} {\bibfnamefont {U.}~\bibnamefont
  {Seifert}},\ }\bibfield  {title} {\bibinfo {title} {Stochastic
  thermodynamics, fluctuation theorems, and molecular machines},\ }\href
  {https://doi.org/10.1088/0034-4885/75/12/126001} {\bibfield  {journal}
  {\bibinfo  {journal} {Rep. Prog. Phys.}\ }\textbf {\bibinfo {volume} {75}},\
  \bibinfo {pages} {126001} (\bibinfo {year} {2012})}\BibitemShut {NoStop}%
\bibitem [{\citenamefont {Ciliberto}(2017)}]{cili17}%
  \BibitemOpen
  \bibfield  {author} {\bibinfo {author} {\bibfnamefont {S.}~\bibnamefont
  {Ciliberto}},\ }\bibfield  {title} {\bibinfo {title} {Experiments in
  stochastic thermodynamics: Short history and perspectives},\ }\href
  {https://doi.org/10.1103/PhysRevX.7.021051} {\bibfield  {journal} {\bibinfo
  {journal} {Phys. Rev. X}\ }\textbf {\bibinfo {volume} {7}},\ \bibinfo {pages}
  {021051} (\bibinfo {year} {2017})}\BibitemShut {NoStop}%
\bibitem [{\citenamefont {Pekola}(2015)}]{peko15}%
  \BibitemOpen
  \bibfield  {author} {\bibinfo {author} {\bibfnamefont {J.~P.}\ \bibnamefont
  {Pekola}},\ }\bibfield  {title} {\bibinfo {title} {Towards quantum
  thermodynamics in electronic circuits},\ }\href
  {https://doi.org/10.1038/nphys3169} {\bibfield  {journal} {\bibinfo
  {journal} {Nature Phys.}\ }\textbf {\bibinfo {volume} {11}},\ \bibinfo
  {pages} {118} (\bibinfo {year} {2015})}\BibitemShut {NoStop}%
\bibitem [{\citenamefont {Liphardt}\ \emph {et~al.}(2002)\citenamefont
  {Liphardt}, \citenamefont {Dumont}, \citenamefont {Smith}, \citenamefont
  {Tinoco~Jr},\ and\ \citenamefont {Bustamante}}]{liph02}%
  \BibitemOpen
  \bibfield  {author} {\bibinfo {author} {\bibfnamefont {J.}~\bibnamefont
  {Liphardt}}, \bibinfo {author} {\bibfnamefont {S.}~\bibnamefont {Dumont}},
  \bibinfo {author} {\bibfnamefont {S.~B.}\ \bibnamefont {Smith}}, \bibinfo
  {author} {\bibfnamefont {I.}~\bibnamefont {Tinoco~Jr}},\ and\ \bibinfo
  {author} {\bibfnamefont {C.}~\bibnamefont {Bustamante}},\ }\bibfield  {title}
  {\bibinfo {title} {Equilibrium information from nonequilibrium measurements
  in an experimental test of {J}arzynski's equality},\ }\href
  {https://doi.org/10.1126/science.1071152} {\bibfield  {journal} {\bibinfo
  {journal} {Science}\ }\textbf {\bibinfo {volume} {296}},\ \bibinfo {pages}
  {1832} (\bibinfo {year} {2002})}\BibitemShut {NoStop}%
\bibitem [{\citenamefont {Collin}\ \emph {et~al.}(2005)\citenamefont {Collin},
  \citenamefont {Ritort}, \citenamefont {{J}arzynski}, \citenamefont {Smith},
  \citenamefont {Tinoco},\ and\ \citenamefont {Bustamante}}]{coll05}%
  \BibitemOpen
  \bibfield  {author} {\bibinfo {author} {\bibfnamefont {D.}~\bibnamefont
  {Collin}}, \bibinfo {author} {\bibfnamefont {F.}~\bibnamefont {Ritort}},
  \bibinfo {author} {\bibfnamefont {C.}~\bibnamefont {{J}arzynski}}, \bibinfo
  {author} {\bibfnamefont {S.}~\bibnamefont {Smith}}, \bibinfo {author}
  {\bibfnamefont {I.}~\bibnamefont {Tinoco}},\ and\ \bibinfo {author}
  {\bibfnamefont {C.}~\bibnamefont {Bustamante}},\ }\bibfield  {title}
  {\bibinfo {title} {Verification of the {C}rooks fluctuation theorem and
  recovery of {RNA} folding free energies},\ }\href
  {http://dx.doi.org/10.1038/nature0406} {\bibfield  {journal} {\bibinfo
  {journal} {Nature}\ }\textbf {\bibinfo {volume} {437}},\ \bibinfo {pages}
  {231} (\bibinfo {year} {2005})}\BibitemShut {NoStop}%
\bibitem [{\citenamefont {Ritort}(2006)}]{rito06}%
  \BibitemOpen
  \bibfield  {author} {\bibinfo {author} {\bibfnamefont {F.}~\bibnamefont
  {Ritort}},\ }\bibfield  {title} {\bibinfo {title} {Single-molecule
  experiments in biological physics: methods and applications},\ }\href
  {https://doi.org/10.1088/0953-8984/18/32/R01} {\bibfield  {journal} {\bibinfo
   {journal} {JPCond}\ }\textbf {\bibinfo {volume} {18}},\ \bibinfo {pages}
  {R531} (\bibinfo {year} {2006})}\BibitemShut {NoStop}%
\bibitem [{\citenamefont {Hayashi}\ \emph {et~al.}(2010)\citenamefont
  {Hayashi}, \citenamefont {Ueno}, \citenamefont {Iino},\ and\ \citenamefont
  {Noji}}]{haya10}%
  \BibitemOpen
  \bibfield  {author} {\bibinfo {author} {\bibfnamefont {K.}~\bibnamefont
  {Hayashi}}, \bibinfo {author} {\bibfnamefont {H.}~\bibnamefont {Ueno}},
  \bibinfo {author} {\bibfnamefont {R.}~\bibnamefont {Iino}},\ and\ \bibinfo
  {author} {\bibfnamefont {H.}~\bibnamefont {Noji}},\ }\bibfield  {title}
  {\bibinfo {title} {Fluctuation theorem applied to {F$_1$}-{ATP}ase},\ }\href
  {https://doi.org/10.1103/PhysRevLett.104.218103} {\bibfield  {journal}
  {\bibinfo  {journal} {PRL}\ }\textbf {\bibinfo {volume} {104}},\ \bibinfo
  {pages} {218103} (\bibinfo {year} {2010})}\BibitemShut {NoStop}%
\bibitem [{\citenamefont {Toyabe}\ \emph {et~al.}(2010)\citenamefont {Toyabe},
  \citenamefont {Okamoto}, \citenamefont {Watanabe-Nakayama}, \citenamefont
  {Taketani}, \citenamefont {Kudo},\ and\ \citenamefont {Muneyuki}}]{toya10}%
  \BibitemOpen
  \bibfield  {author} {\bibinfo {author} {\bibfnamefont {S.}~\bibnamefont
  {Toyabe}}, \bibinfo {author} {\bibfnamefont {T.}~\bibnamefont {Okamoto}},
  \bibinfo {author} {\bibfnamefont {T.}~\bibnamefont {Watanabe-Nakayama}},
  \bibinfo {author} {\bibfnamefont {H.}~\bibnamefont {Taketani}}, \bibinfo
  {author} {\bibfnamefont {S.}~\bibnamefont {Kudo}},\ and\ \bibinfo {author}
  {\bibfnamefont {E.}~\bibnamefont {Muneyuki}},\ }\bibfield  {title} {\bibinfo
  {title} {Nonequilibrium energetics of a single {F$_1$}-{ATP}ase molecule},\
  }\href {https://doi.org/10.1103/PhysRevLett.104.198103} {\bibfield  {journal}
  {\bibinfo  {journal} {PRL}\ }\textbf {\bibinfo {volume} {104}},\ \bibinfo
  {pages} {198103} (\bibinfo {year} {2010})}\BibitemShut {NoStop}%
\bibitem [{\citenamefont {Li}\ and\ \citenamefont {Toyabe}(2020)}]{li20}%
  \BibitemOpen
  \bibfield  {author} {\bibinfo {author} {\bibfnamefont {C.-B.}\ \bibnamefont
  {Li}}\ and\ \bibinfo {author} {\bibfnamefont {S.}~\bibnamefont {Toyabe}},\
  }\bibfield  {title} {\bibinfo {title} {Efficiencies of molecular motors: a
  comprehensible overview},\ }\href
  {https://doi.org/10.1007/s12551-020-00672-x} {\bibfield  {journal} {\bibinfo
  {journal} {Biophysical Reviews}\ }\textbf {\bibinfo {volume} {12}},\ \bibinfo
  {pages} {419–423} (\bibinfo {year} {2020})}\BibitemShut {NoStop}%
\bibitem [{\citenamefont {Barato}\ and\ \citenamefont
  {Seifert}(2015)}]{bara15}%
  \BibitemOpen
  \bibfield  {author} {\bibinfo {author} {\bibfnamefont {A.~C.}\ \bibnamefont
  {Barato}}\ and\ \bibinfo {author} {\bibfnamefont {U.}~\bibnamefont
  {Seifert}},\ }\bibfield  {title} {\bibinfo {title} {Thermodynamic uncertainty
  relation for biomolecular processes},\ }\href
  {https://doi.org/10.1103/PhysRevLett.114.158101} {\bibfield  {journal}
  {\bibinfo  {journal} {PRL}\ }\textbf {\bibinfo {volume} {114}},\ \bibinfo
  {pages} {158101} (\bibinfo {year} {2015})}\BibitemShut {NoStop}%
\bibitem [{\citenamefont {Gingrich}\ \emph {et~al.}(2016)\citenamefont
  {Gingrich}, \citenamefont {Horowitz}, \citenamefont {Perunov},\ and\
  \citenamefont {England}}]{ging16}%
  \BibitemOpen
  \bibfield  {author} {\bibinfo {author} {\bibfnamefont {T.~R.}\ \bibnamefont
  {Gingrich}}, \bibinfo {author} {\bibfnamefont {J.~M.}\ \bibnamefont
  {Horowitz}}, \bibinfo {author} {\bibfnamefont {N.}~\bibnamefont {Perunov}},\
  and\ \bibinfo {author} {\bibfnamefont {J.~L.}\ \bibnamefont {England}},\
  }\bibfield  {title} {\bibinfo {title} {Dissipation bounds all steady-state
  current fluctuations},\ }\href
  {https://doi.org/10.1103/PhysRevLett.116.120601} {\bibfield  {journal}
  {\bibinfo  {journal} {PRL}\ }\textbf {\bibinfo {volume} {116}},\ \bibinfo
  {pages} {120601} (\bibinfo {year} {2016})}\BibitemShut {NoStop}%
\bibitem [{\citenamefont {Horowitz}\ and\ \citenamefont
  {Gingrich}(2020)}]{horo20}%
  \BibitemOpen
  \bibfield  {author} {\bibinfo {author} {\bibfnamefont {J.~M.}\ \bibnamefont
  {Horowitz}}\ and\ \bibinfo {author} {\bibfnamefont {T.~R.}\ \bibnamefont
  {Gingrich}},\ }\bibfield  {title} {\bibinfo {title} {Thermodynamic
  uncertainty relations constrain non-equilibrium fluctuations},\ }\href
  {https://doi.org/https://doi.org/10.1038/s41567-019-0702-6} {\bibfield
  {journal} {\bibinfo  {journal} {Nat. Phys.}\ }\textbf {\bibinfo {volume}
  {16}},\ \bibinfo {pages} {15} (\bibinfo {year} {2020})}\BibitemShut {NoStop}%
\bibitem [{\citenamefont {Cao}\ \emph {et~al.}(2015)\citenamefont {Cao},
  \citenamefont {Wang}, \citenamefont {Ouyang},\ and\ \citenamefont
  {Tu}}]{cao15}%
  \BibitemOpen
  \bibfield  {author} {\bibinfo {author} {\bibfnamefont {Y.}~\bibnamefont
  {Cao}}, \bibinfo {author} {\bibfnamefont {H.}~\bibnamefont {Wang}}, \bibinfo
  {author} {\bibfnamefont {Q.}~\bibnamefont {Ouyang}},\ and\ \bibinfo {author}
  {\bibfnamefont {Y.}~\bibnamefont {Tu}},\ }\bibfield  {title} {\bibinfo
  {title} {The free-energy cost of accurate biochemical oscillations},\ }\href
  {https://doi.org/10.1038/nphys3412} {\bibfield  {journal} {\bibinfo
  {journal} {Nature Phys.}\ }\textbf {\bibinfo {volume} {11}},\ \bibinfo
  {pages} {772} (\bibinfo {year} {2015})}\BibitemShut {NoStop}%
\bibitem [{\citenamefont {Oberreiter}\ \emph {et~al.}(2022)\citenamefont
  {Oberreiter}, \citenamefont {Seifert},\ and\ \citenamefont
  {Barato}}]{ober22}%
  \BibitemOpen
  \bibfield  {author} {\bibinfo {author} {\bibfnamefont {L.}~\bibnamefont
  {Oberreiter}}, \bibinfo {author} {\bibfnamefont {U.}~\bibnamefont
  {Seifert}},\ and\ \bibinfo {author} {\bibfnamefont {A.~C.}\ \bibnamefont
  {Barato}},\ }\bibfield  {title} {\bibinfo {title} {Universal minimal cost of
  coherent biochemical oscillations},\ }\href
  {https://doi.org/10.1103/PhysRevE.106.014106} {\bibfield  {journal} {\bibinfo
   {journal} {Phys. Rev. E}\ }\textbf {\bibinfo {volume} {106}},\ \bibinfo
  {pages} {014106} (\bibinfo {year} {2022})}\BibitemShut {NoStop}%
\bibitem [{\citenamefont {Ohga}\ \emph {et~al.}(2023)\citenamefont {Ohga},
  \citenamefont {Ito},\ and\ \citenamefont {Kolchinsky}}]{ohga23}%
  \BibitemOpen
  \bibfield  {author} {\bibinfo {author} {\bibfnamefont {N.}~\bibnamefont
  {Ohga}}, \bibinfo {author} {\bibfnamefont {S.}~\bibnamefont {Ito}},\ and\
  \bibinfo {author} {\bibfnamefont {A.}~\bibnamefont {Kolchinsky}},\ }\bibfield
   {title} {\bibinfo {title} {Thermodynamic bound on the asymmetry of
  cross-correlations},\ }\href {https://doi.org/10.1103/PhysRevLett.131.077101}
  {\bibfield  {journal} {\bibinfo  {journal} {Phys. Rev. Lett.}\ }\textbf
  {\bibinfo {volume} {131}},\ \bibinfo {pages} {077101} (\bibinfo {year}
  {2023})}\BibitemShut {NoStop}%
\bibitem [{\citenamefont {Aurell}\ \emph {et~al.}(2011)\citenamefont {Aurell},
  \citenamefont {Mej{\'i}a-Monasterio},\ and\ \citenamefont
  {Muratore-Ginanneschi}}]{aure11}%
  \BibitemOpen
  \bibfield  {author} {\bibinfo {author} {\bibfnamefont {E.}~\bibnamefont
  {Aurell}}, \bibinfo {author} {\bibfnamefont {C.}~\bibnamefont
  {Mej{\'i}a-Monasterio}},\ and\ \bibinfo {author} {\bibfnamefont
  {P.}~\bibnamefont {Muratore-Ginanneschi}},\ }\bibfield  {title} {\bibinfo
  {title} {Optimal protocols and optimal transport in stochastic
  thermodynamics},\ }\href {https://doi.org/10.1103/PhysRevLett.106.250601}
  {\bibfield  {journal} {\bibinfo  {journal} {PRL}\ }\textbf {\bibinfo {volume}
  {106}},\ \bibinfo {pages} {250601} (\bibinfo {year} {2011})}\BibitemShut
  {NoStop}%
\bibitem [{\citenamefont {Shiraishi}\ \emph {et~al.}(2018)\citenamefont
  {Shiraishi}, \citenamefont {Funo},\ and\ \citenamefont {Saito}}]{shir18}%
  \BibitemOpen
  \bibfield  {author} {\bibinfo {author} {\bibfnamefont {N.}~\bibnamefont
  {Shiraishi}}, \bibinfo {author} {\bibfnamefont {K.}~\bibnamefont {Funo}},\
  and\ \bibinfo {author} {\bibfnamefont {K.}~\bibnamefont {Saito}},\ }\bibfield
   {title} {\bibinfo {title} {Speed limit for classical stochastic processes},\
  }\href {https://doi.org/10.1103/PhysRevLett.121.070601} {\bibfield  {journal}
  {\bibinfo  {journal} {Phys. Rev. Lett.}\ }\textbf {\bibinfo {volume} {121}},\
  \bibinfo {pages} {070601} (\bibinfo {year} {2018})}\BibitemShut {NoStop}%
\bibitem [{\citenamefont {Ito}\ and\ \citenamefont {Dechant}(2020)}]{ito20}%
  \BibitemOpen
  \bibfield  {author} {\bibinfo {author} {\bibfnamefont {S.}~\bibnamefont
  {Ito}}\ and\ \bibinfo {author} {\bibfnamefont {A.}~\bibnamefont {Dechant}},\
  }\bibfield  {title} {\bibinfo {title} {Stochastic time evolution, information
  geometry, and the {Cram\'er}-{Rao} bound},\ }\href
  {https://doi.org/10.1103/PhysRevX.10.021056} {\bibfield  {journal} {\bibinfo
  {journal} {Phys. Rev. X}\ }\textbf {\bibinfo {volume} {10}},\ \bibinfo
  {pages} {021056} (\bibinfo {year} {2020})}\BibitemShut {NoStop}%
\bibitem [{\citenamefont {Van~Vu}\ and\ \citenamefont {Saito}(2023)}]{vu23}%
  \BibitemOpen
  \bibfield  {author} {\bibinfo {author} {\bibfnamefont {T.}~\bibnamefont
  {Van~Vu}}\ and\ \bibinfo {author} {\bibfnamefont {K.}~\bibnamefont {Saito}},\
  }\bibfield  {title} {\bibinfo {title} {Thermodynamic unification of optimal
  transport: Thermodynamic uncertainty relation, minimum dissipation, and
  thermodynamic speed limits},\ }\href
  {https://doi.org/10.1103/PhysRevX.13.011013} {\bibfield  {journal} {\bibinfo
  {journal} {Phys. Rev. X}\ }\textbf {\bibinfo {volume} {13}},\ \bibinfo
  {pages} {011013} (\bibinfo {year} {2023})}\BibitemShut {NoStop}%
\bibitem [{\citenamefont {Berezhkovskii}\ and\ \citenamefont
  {Makarov}(2019)}]{bere19}%
  \BibitemOpen
  \bibfield  {author} {\bibinfo {author} {\bibfnamefont {A.~M.}\ \bibnamefont
  {Berezhkovskii}}\ and\ \bibinfo {author} {\bibfnamefont {D.~E.}\ \bibnamefont
  {Makarov}},\ }\bibfield  {title} {\bibinfo {title} {{On the forward/backward
  symmetry of transition path time distributions in nonequilibrium systems}},\
  }\href {https://doi.org/10.1063/1.5109293} {\bibfield  {journal} {\bibinfo
  {journal} {The Journal of Chemical Physics}\ }\textbf {\bibinfo {volume}
  {151}},\ \bibinfo {pages} {065102} (\bibinfo {year} {2019})}\BibitemShut
  {NoStop}%
\bibitem [{\citenamefont {Gladrow}\ \emph {et~al.}(2019)\citenamefont
  {Gladrow}, \citenamefont {Ribezzi-Crivellari}, \citenamefont {Ritort},\ and\
  \citenamefont {Keyser}}]{glad19}%
  \BibitemOpen
  \bibfield  {author} {\bibinfo {author} {\bibfnamefont {J.}~\bibnamefont
  {Gladrow}}, \bibinfo {author} {\bibfnamefont {M.}~\bibnamefont
  {Ribezzi-Crivellari}}, \bibinfo {author} {\bibfnamefont {F.}~\bibnamefont
  {Ritort}},\ and\ \bibinfo {author} {\bibfnamefont {U.~F.}\ \bibnamefont
  {Keyser}},\ }\bibfield  {title} {\bibinfo {title} {Experimental evidence of
  symmetry breaking of transition-path times},\ }\bibfield  {journal} {\bibinfo
   {journal} {Nature Communications}\ }\textbf {\bibinfo {volume} {10}},\ \href
  {https://doi.org/10.1038/s41467-018-07873-9} {10.1038/s41467-018-07873-9}
  (\bibinfo {year} {2019})\BibitemShut {NoStop}%
\bibitem [{\citenamefont {Skinner}\ and\ \citenamefont
  {Dunkel}(2021)}]{skin21a}%
  \BibitemOpen
  \bibfield  {author} {\bibinfo {author} {\bibfnamefont {D.~J.}\ \bibnamefont
  {Skinner}}\ and\ \bibinfo {author} {\bibfnamefont {J.}~\bibnamefont
  {Dunkel}},\ }\bibfield  {title} {\bibinfo {title} {Estimating entropy
  production from waiting time distributions},\ }\href
  {https://link.aps.org/doi/10.1103/PhysRevLett.127.198101} {\bibfield
  {journal} {\bibinfo  {journal} {PRL}\ }\textbf {\bibinfo {volume} {127}},\
  \bibinfo {pages} {198101} (\bibinfo {year} {2021})}\BibitemShut {NoStop}%
\bibitem [{\citenamefont {Fodor}\ \emph {et~al.}(2022)\citenamefont {Fodor},
  \citenamefont {Jack},\ and\ \citenamefont {Cates}}]{fodo22}%
  \BibitemOpen
  \bibfield  {author} {\bibinfo {author} {\bibfnamefont {{\'E}.}~\bibnamefont
  {Fodor}}, \bibinfo {author} {\bibfnamefont {R.~L.}\ \bibnamefont {Jack}},\
  and\ \bibinfo {author} {\bibfnamefont {M.~E.}\ \bibnamefont {Cates}},\
  }\bibfield  {title} {\bibinfo {title} {Irreversibility and biased ensembles
  in active matter: Insights from stochastic thermodynamics},\ }\href
  {https://doi.org/10.1146/annurev-conmatphys-031720-032419} {\bibfield
  {journal} {\bibinfo  {journal} {Annual Review of Condensed Matter Physics}\
  }\textbf {\bibinfo {volume} {13}},\ \bibinfo {pages} {215} (\bibinfo {year}
  {2022})},\ \Eprint
  {https://arxiv.org/abs/https://doi.org/10.1146/annurev-conmatphys-031720-032419}
  {https://doi.org/10.1146/annurev-conmatphys-031720-032419} \BibitemShut
  {NoStop}%
\bibitem [{\citenamefont {Hwang}\ and\ \citenamefont {Hyeon}(2018)}]{hwan18}%
  \BibitemOpen
  \bibfield  {author} {\bibinfo {author} {\bibfnamefont {W.}~\bibnamefont
  {Hwang}}\ and\ \bibinfo {author} {\bibfnamefont {C.}~\bibnamefont {Hyeon}},\
  }\bibfield  {title} {\bibinfo {title} {Energetic costs, precision, and
  transport efficiency of molecular motors},\ }\href
  {https://doi.org/10.1021/acs.jpclett.7b03197} {\bibfield  {journal} {\bibinfo
   {journal} {J. Phys. Chem. Lett.}\ }\textbf {\bibinfo {volume} {9}},\
  \bibinfo {pages} {513} (\bibinfo {year} {2018})}\BibitemShut {NoStop}%
\bibitem [{\citenamefont {Li}\ \emph {et~al.}(2019)\citenamefont {Li},
  \citenamefont {Horowitz}, \citenamefont {Gingrich},\ and\ \citenamefont
  {Fakhri}}]{li19}%
  \BibitemOpen
  \bibfield  {author} {\bibinfo {author} {\bibfnamefont {J.}~\bibnamefont
  {Li}}, \bibinfo {author} {\bibfnamefont {J.~M.}\ \bibnamefont {Horowitz}},
  \bibinfo {author} {\bibfnamefont {T.~R.}\ \bibnamefont {Gingrich}},\ and\
  \bibinfo {author} {\bibfnamefont {N.}~\bibnamefont {Fakhri}},\ }\bibfield
  {title} {\bibinfo {title} {Quantifying dissipation using fluctuating
  currents},\ }\href
  {https://doi.org/https://doi.org/10.1038/s41467-019-09631-x} {\bibfield
  {journal} {\bibinfo  {journal} {Nat. Commun.}\ }\textbf {\bibinfo {volume}
  {10}},\ \bibinfo {pages} {1666} (\bibinfo {year} {2019})}\BibitemShut
  {NoStop}%
\bibitem [{\citenamefont {Van~Vu}\ \emph {et~al.}(2020)\citenamefont {Van~Vu},
  \citenamefont {Vo},\ and\ \citenamefont {Hasegawa}}]{vu20}%
  \BibitemOpen
  \bibfield  {author} {\bibinfo {author} {\bibfnamefont {T.}~\bibnamefont
  {Van~Vu}}, \bibinfo {author} {\bibfnamefont {V.~T.}\ \bibnamefont {Vo}},\
  and\ \bibinfo {author} {\bibfnamefont {Y.}~\bibnamefont {Hasegawa}},\
  }\bibfield  {title} {\bibinfo {title} {Entropy production estimation with
  optimal current},\ }\href {https://doi.org/10.1103/PhysRevE.101.042138}
  {\bibfield  {journal} {\bibinfo  {journal} {Physical Review E}\ }\textbf
  {\bibinfo {volume} {101}},\ \bibinfo {pages} {042138} (\bibinfo {year}
  {2020})}\BibitemShut {NoStop}%
\bibitem [{\citenamefont {Koyuk}\ and\ \citenamefont {Seifert}(2020)}]{koyu20}%
  \BibitemOpen
  \bibfield  {author} {\bibinfo {author} {\bibfnamefont {T.}~\bibnamefont
  {Koyuk}}\ and\ \bibinfo {author} {\bibfnamefont {U.}~\bibnamefont
  {Seifert}},\ }\bibfield  {title} {\bibinfo {title} {Thermodynamic uncertainty
  relation for time-dependent driving},\ }\href
  {https://doi.org/10.1103/PhysRevLett.125.260604} {\bibfield  {journal}
  {\bibinfo  {journal} {PRL}\ }\textbf {\bibinfo {volume} {125}},\ \bibinfo
  {pages} {260604} (\bibinfo {year} {2020})}\BibitemShut {NoStop}%
\bibitem [{\citenamefont {Dechant}\ and\ \citenamefont {Sasa}(2021)}]{dech21}%
  \BibitemOpen
  \bibfield  {author} {\bibinfo {author} {\bibfnamefont {A.}~\bibnamefont
  {Dechant}}\ and\ \bibinfo {author} {\bibfnamefont {S.~I.}\ \bibnamefont
  {Sasa}},\ }\bibfield  {title} {\bibinfo {title} {Improving thermodynamic
  bounds using correlations},\ }\href
  {https://doi.org/https://doi.org/10.1103/PhysRevX.11.041061} {\bibfield
  {journal} {\bibinfo  {journal} {Phys. Rev. X}\ }\textbf {\bibinfo {volume}
  {11}},\ \bibinfo {pages} {041061} (\bibinfo {year} {2021})}\BibitemShut
  {NoStop}%
\bibitem [{\citenamefont {Dechant}\ \emph {et~al.}(2023)\citenamefont
  {Dechant}, \citenamefont {Garnier-Brun},\ and\ \citenamefont
  {Sasa}}]{dech23}%
  \BibitemOpen
  \bibfield  {author} {\bibinfo {author} {\bibfnamefont {A.}~\bibnamefont
  {Dechant}}, \bibinfo {author} {\bibfnamefont {J.}~\bibnamefont
  {Garnier-Brun}},\ and\ \bibinfo {author} {\bibfnamefont {S.-i.}\ \bibnamefont
  {Sasa}},\ }\bibfield  {title} {\bibinfo {title} {Thermodynamic bounds on
  correlation times},\ }\href {https://doi.org/10.1103/PhysRevLett.131.167101}
  {\bibfield  {journal} {\bibinfo  {journal} {Phys. Rev. Lett.}\ }\textbf
  {\bibinfo {volume} {131}},\ \bibinfo {pages} {167101} (\bibinfo {year}
  {2023})}\BibitemShut {NoStop}%
\bibitem [{\citenamefont {Pietzonka}\ and\ \citenamefont
  {Coghi}(2023)}]{piet23}%
  \BibitemOpen
  \bibfield  {author} {\bibinfo {author} {\bibfnamefont {P.}~\bibnamefont
  {Pietzonka}}\ and\ \bibinfo {author} {\bibfnamefont {F.}~\bibnamefont
  {Coghi}},\ }\bibfield  {title} {\bibinfo {title} {Thermodynamic cost for
  precision of general counting observables},\ }\href@noop {} {\bibfield
  {journal} {\bibinfo  {journal} {arXiv:2305.15392}\ } (\bibinfo {year}
  {2023})}\BibitemShut {NoStop}%
\bibitem [{\citenamefont {Neri}\ and\ \citenamefont
  {Polettini}(2023)}]{neri23}%
  \BibitemOpen
  \bibfield  {author} {\bibinfo {author} {\bibfnamefont {I.}~\bibnamefont
  {Neri}}\ and\ \bibinfo {author} {\bibfnamefont {M.}~\bibnamefont
  {Polettini}},\ }\bibfield  {title} {\bibinfo {title} {{Extreme value
  statistics of edge currents in Markov jump processes and their use for
  entropy production estimation}},\ }\href
  {https://doi.org/10.21468/SciPostPhys.14.5.131} {\bibfield  {journal}
  {\bibinfo  {journal} {SciPost Phys.}\ }\textbf {\bibinfo {volume} {14}},\
  \bibinfo {pages} {131} (\bibinfo {year} {2023})}\BibitemShut {NoStop}%
\bibitem [{\citenamefont {Cover}\ and\ \citenamefont {Thomas}(2006)}]{cove06}%
  \BibitemOpen
  \bibfield  {author} {\bibinfo {author} {\bibfnamefont {T.~M.}\ \bibnamefont
  {Cover}}\ and\ \bibinfo {author} {\bibfnamefont {J.~A.}\ \bibnamefont
  {Thomas}},\ }\href@noop {} {\emph {\bibinfo {title} {Elements of information
  theory}}},\ Telecommunications and signal processing\ (\bibinfo  {publisher}
  {Wiley},\ \bibinfo {address} {Hoboken, NJ, and Canada},\ \bibinfo {year}
  {2006})\BibitemShut {NoStop}%
\bibitem [{\citenamefont {Kawai}\ \emph {et~al.}(2007)\citenamefont {Kawai},
  \citenamefont {Parrondo},\ and\ \citenamefont {{van den Broeck}}}]{kawa07}%
  \BibitemOpen
  \bibfield  {author} {\bibinfo {author} {\bibfnamefont {R.}~\bibnamefont
  {Kawai}}, \bibinfo {author} {\bibfnamefont {J.~M.~R.}\ \bibnamefont
  {Parrondo}},\ and\ \bibinfo {author} {\bibfnamefont {C.}~\bibnamefont {{van
  den Broeck}}},\ }\bibfield  {title} {\bibinfo {title} {Dissipation: The
  phase-space perspective},\ }\href
  {https://doi.org/10.1103/PhysRevLett.98.080602} {\bibfield  {journal}
  {\bibinfo  {journal} {PRL}\ }\textbf {\bibinfo {volume} {98}},\ \bibinfo
  {pages} {080602} (\bibinfo {year} {2007})}\BibitemShut {NoStop}%
\bibitem [{\citenamefont {Roldan}\ and\ \citenamefont
  {Parrondo}(2010)}]{rold10}%
  \BibitemOpen
  \bibfield  {author} {\bibinfo {author} {\bibfnamefont {E.}~\bibnamefont
  {Roldan}}\ and\ \bibinfo {author} {\bibfnamefont {J.~M.~R.}\ \bibnamefont
  {Parrondo}},\ }\bibfield  {title} {\bibinfo {title} {Estimating dissipation
  from single stationary trajectories},\ }\href
  {https://doi.org/10.1103/PhysRevLett.105.150607} {\bibfield  {journal}
  {\bibinfo  {journal} {PRL}\ }\textbf {\bibinfo {volume} {105}},\ \bibinfo
  {pages} {150607} (\bibinfo {year} {2010})}\BibitemShut {NoStop}%
\bibitem [{\citenamefont {Shiraishi}\ and\ \citenamefont
  {Sagawa}(2015)}]{shir14}%
  \BibitemOpen
  \bibfield  {author} {\bibinfo {author} {\bibfnamefont {N.}~\bibnamefont
  {Shiraishi}}\ and\ \bibinfo {author} {\bibfnamefont {T.}~\bibnamefont
  {Sagawa}},\ }\bibfield  {title} {\bibinfo {title} {Fluctuation theorem for
  partially masked nonequilibrium dynamics},\ }\href
  {https://doi.org/https://doi.org/10.1103/PhysRevE.91.012130} {\bibfield
  {journal} {\bibinfo  {journal} {PRE}\ }\textbf {\bibinfo {volume} {91}},\
  \bibinfo {pages} {012130} (\bibinfo {year} {2015})}\BibitemShut {NoStop}%
\bibitem [{\citenamefont {Bisker}\ \emph {et~al.}(2017)\citenamefont {Bisker},
  \citenamefont {Polettini}, \citenamefont {Gingrich},\ and\ \citenamefont
  {Horowitz}}]{bisk17}%
  \BibitemOpen
  \bibfield  {author} {\bibinfo {author} {\bibfnamefont {G.}~\bibnamefont
  {Bisker}}, \bibinfo {author} {\bibfnamefont {M.}~\bibnamefont {Polettini}},
  \bibinfo {author} {\bibfnamefont {T.~R.}\ \bibnamefont {Gingrich}},\ and\
  \bibinfo {author} {\bibfnamefont {J.~M.}\ \bibnamefont {Horowitz}},\
  }\bibfield  {title} {\bibinfo {title} {Hierarchical bounds on entropy
  production inferred from partial information},\ }\href
  {https://doi.org/10.1088/1742-5468/aa8c0d} {\bibfield  {journal} {\bibinfo
  {journal} {J. Stat. Mech. Theor. Exp.}\ }\textbf {\bibinfo {volume} {2017}},\
  \bibinfo {pages} {093210} (\bibinfo {year} {2017})}\BibitemShut {NoStop}%
\bibitem [{\citenamefont {Mart{\'i}nez}\ \emph {et~al.}(2019)\citenamefont
  {Mart{\'i}nez}, \citenamefont {Bisker}, \citenamefont {Horowitz},\ and\
  \citenamefont {Parrondo}}]{mart19}%
  \BibitemOpen
  \bibfield  {author} {\bibinfo {author} {\bibfnamefont {I.~A.}\ \bibnamefont
  {Mart{\'i}nez}}, \bibinfo {author} {\bibfnamefont {G.}~\bibnamefont
  {Bisker}}, \bibinfo {author} {\bibfnamefont {J.~M.}\ \bibnamefont
  {Horowitz}},\ and\ \bibinfo {author} {\bibfnamefont {J.~M.~R.}\ \bibnamefont
  {Parrondo}},\ }\bibfield  {title} {\bibinfo {title} {Inferring broken
  detailed balance in the absence of observable currents},\ }\href
  {https://doi.org/10.1038/s41467-019-11051-w} {\bibfield  {journal} {\bibinfo
  {journal} {Nature Communications}\ }\textbf {\bibinfo {volume} {10}},\
  \bibinfo {pages} {3542} (\bibinfo {year} {2019})}\BibitemShut {NoStop}%
\bibitem [{\citenamefont {Hartich}\ and\ \citenamefont
  {Godec}(2021)}]{hart21a}%
  \BibitemOpen
  \bibfield  {author} {\bibinfo {author} {\bibfnamefont {D.}~\bibnamefont
  {Hartich}}\ and\ \bibinfo {author} {\bibfnamefont {A.}~\bibnamefont
  {Godec}},\ }\bibfield  {title} {\bibinfo {title} {Emergent memory and kinetic
  hysteresis in strongly driven networks},\ }\href
  {https://doi.org/https://doi.org/10.1103/PhysRevX.11.041047} {\bibfield
  {journal} {\bibinfo  {journal} {Phys. Rev. X}\ }\textbf {\bibinfo {volume}
  {11}},\ \bibinfo {pages} {041047} (\bibinfo {year} {2021})}\BibitemShut
  {NoStop}%
\bibitem [{\citenamefont {{van der Meer}}\ \emph {et~al.}(2022)\citenamefont
  {{van der Meer}}, \citenamefont {Ertel},\ and\ \citenamefont
  {Seifert}}]{vdm22}%
  \BibitemOpen
  \bibfield  {author} {\bibinfo {author} {\bibfnamefont {J.}~\bibnamefont {{van
  der Meer}}}, \bibinfo {author} {\bibfnamefont {B.}~\bibnamefont {Ertel}},\
  and\ \bibinfo {author} {\bibfnamefont {U.}~\bibnamefont {Seifert}},\
  }\bibfield  {title} {\bibinfo {title} {Thermodynamic inference in partially
  accessible markov networks: A unifying perspective from transition-based
  waiting time distributions},\ }\href
  {https://doi.org/https://doi.org/10.1103/PhysRevX.12.031025} {\bibfield
  {journal} {\bibinfo  {journal} {Phys. Rev. X}\ }\textbf {\bibinfo {volume}
  {12}},\ \bibinfo {pages} {031025} (\bibinfo {year} {2022})}\BibitemShut
  {NoStop}%
\bibitem [{\citenamefont {Harunari}\ \emph {et~al.}(2022)\citenamefont
  {Harunari}, \citenamefont {Dutta}, \citenamefont {Polettini},\ and\
  \citenamefont {Roldan}}]{haru22}%
  \BibitemOpen
  \bibfield  {author} {\bibinfo {author} {\bibfnamefont {P.}~\bibnamefont
  {Harunari}}, \bibinfo {author} {\bibfnamefont {A.}~\bibnamefont {Dutta}},
  \bibinfo {author} {\bibfnamefont {M.}~\bibnamefont {Polettini}},\ and\
  \bibinfo {author} {\bibfnamefont {E.}~\bibnamefont {Roldan}},\ }\bibfield
  {title} {\bibinfo {title} {What to learn from a few visible transitions’
  statistics?},\ }\href
  {https://doi.org/https://doi.org/10.1103/PhysRevX.12.041026} {\bibfield
  {journal} {\bibinfo  {journal} {Phys. Rev. X}\ }\textbf {\bibinfo {volume}
  {12}},\ \bibinfo {pages} {041026} (\bibinfo {year} {2022})}\BibitemShut
  {NoStop}%
\bibitem [{\citenamefont {van~der Meer}\ \emph {et~al.}(2023)\citenamefont
  {van~der Meer}, \citenamefont {Deg\"unther},\ and\ \citenamefont
  {Seifert}}]{vdm23}%
  \BibitemOpen
  \bibfield  {author} {\bibinfo {author} {\bibfnamefont {J.}~\bibnamefont
  {van~der Meer}}, \bibinfo {author} {\bibfnamefont {J.}~\bibnamefont
  {Deg\"unther}},\ and\ \bibinfo {author} {\bibfnamefont {U.}~\bibnamefont
  {Seifert}},\ }\bibfield  {title} {\bibinfo {title} {Time-resolved statistics
  of snippets as general framework for model-free entropy estimators},\ }\href
  {https://doi.org/10.1103/PhysRevLett.130.257101} {\bibfield  {journal}
  {\bibinfo  {journal} {Phys. Rev. Lett.}\ }\textbf {\bibinfo {volume} {130}},\
  \bibinfo {pages} {257101} (\bibinfo {year} {2023})}\BibitemShut {NoStop}%
\bibitem [{\citenamefont {Deg\"unther}\ \emph {et~al.}(2024)\citenamefont
  {Deg\"unther}, \citenamefont {van~der Meer},\ and\ \citenamefont
  {Seifert}}]{degu23}%
  \BibitemOpen
  \bibfield  {author} {\bibinfo {author} {\bibfnamefont {J.}~\bibnamefont
  {Deg\"unther}}, \bibinfo {author} {\bibfnamefont {J.}~\bibnamefont {van~der
  Meer}},\ and\ \bibinfo {author} {\bibfnamefont {U.}~\bibnamefont {Seifert}},\
  }\bibfield  {title} {\bibinfo {title} {Fluctuating entropy production on the
  coarse-grained level: Inference and localization of irreversibility},\ }\href
  {https://doi.org/10.1103/PhysRevResearch.6.023175} {\bibfield  {journal}
  {\bibinfo  {journal} {Phys. Rev. Res.}\ }\textbf {\bibinfo {volume} {6}},\
  \bibinfo {pages} {023175} (\bibinfo {year} {2024})}\BibitemShut {NoStop}%
\bibitem [{\citenamefont {Esposito}(2012)}]{espo12}%
  \BibitemOpen
  \bibfield  {author} {\bibinfo {author} {\bibfnamefont {M.}~\bibnamefont
  {Esposito}},\ }\bibfield  {title} {\bibinfo {title} {Stochastic
  thermodynamics under coarse-graining},\ }\href
  {https://doi.org/10.1103/PhysRevE.85.041125} {\bibfield  {journal} {\bibinfo
  {journal} {PRE}\ }\textbf {\bibinfo {volume} {85}},\ \bibinfo {pages}
  {041125} (\bibinfo {year} {2012})}\BibitemShut {NoStop}%
\bibitem [{\citenamefont {Seifert}(2019)}]{seif19}%
  \BibitemOpen
  \bibfield  {author} {\bibinfo {author} {\bibfnamefont {U.}~\bibnamefont
  {Seifert}},\ }\bibfield  {title} {\bibinfo {title} {From stochastic
  thermodynamics to thermodynamic inference},\ }\href
  {https://doi.org/10.1146/annurev-conmatphys-031218-013554} {\bibfield
  {journal} {\bibinfo  {journal} {Ann. Rev. Cond. Mat. Phys.}\ }\textbf
  {\bibinfo {volume} {10}},\ \bibinfo {pages} {171} (\bibinfo {year}
  {2019})}\BibitemShut {NoStop}%
\bibitem [{\citenamefont {Hartich}\ and\ \citenamefont {Godec}(2023)}]{hart23}%
  \BibitemOpen
  \bibfield  {author} {\bibinfo {author} {\bibfnamefont {D.}~\bibnamefont
  {Hartich}}\ and\ \bibinfo {author} {\bibfnamefont {A.}~\bibnamefont
  {Godec}},\ }\bibfield  {title} {\bibinfo {title} {Violation of local detailed
  balance upon lumping despite a clear timescale separation},\ }\href
  {https://doi.org/10.1103/PhysRevResearch.5.L032017} {\bibfield  {journal}
  {\bibinfo  {journal} {Phys. Rev. Res.}\ }\textbf {\bibinfo {volume} {5}},\
  \bibinfo {pages} {L032017} (\bibinfo {year} {2023})}\BibitemShut {NoStop}%
\bibitem [{\citenamefont {Blom}\ \emph {et~al.}(2024)\citenamefont {Blom},
  \citenamefont {Song}, \citenamefont {Vouga}, \citenamefont {Godec},\ and\
  \citenamefont {Makarov}}]{blom23}%
  \BibitemOpen
  \bibfield  {author} {\bibinfo {author} {\bibfnamefont {K.}~\bibnamefont
  {Blom}}, \bibinfo {author} {\bibfnamefont {K.}~\bibnamefont {Song}}, \bibinfo
  {author} {\bibfnamefont {E.}~\bibnamefont {Vouga}}, \bibinfo {author}
  {\bibfnamefont {A.}~\bibnamefont {Godec}},\ and\ \bibinfo {author}
  {\bibfnamefont {D.~E.}\ \bibnamefont {Makarov}},\ }\bibfield  {title}
  {\bibinfo {title} {Milestoning estimators of dissipation in systems observed
  at a coarse resolution},\ }\href {https://doi.org/10.1073/pnas.2318333121}
  {\bibfield  {journal} {\bibinfo  {journal} {Proceedings of the National
  Academy of Sciences}\ }\textbf {\bibinfo {volume} {121}},\ \bibinfo {pages}
  {e2318333121} (\bibinfo {year} {2024})}\BibitemShut {NoStop}%
\bibitem [{\citenamefont {Godec}\ and\ \citenamefont {Makarov}(2023)}]{gode23}%
  \BibitemOpen
  \bibfield  {author} {\bibinfo {author} {\bibfnamefont {A.}~\bibnamefont
  {Godec}}\ and\ \bibinfo {author} {\bibfnamefont {D.~E.}\ \bibnamefont
  {Makarov}},\ }\bibfield  {title} {{\selectlanguage {en}\bibinfo {title}
  {Challenges in {Inferring} the {Directionality} of {Active} {Molecular}
  {Processes} from {Single}-{Molecule} {Fluorescence} {Resonance} {Energy}
  {Transfer} {Trajectories}}},\ }\href
  {https://doi.org/10.1021/acs.jpclett.2c03244} {\bibfield  {journal} {\bibinfo
   {journal} {The Journal of Physical Chemistry Letters}\ }\textbf {\bibinfo
  {volume} {14}},\ \bibinfo {pages} {49} (\bibinfo {year} {2023})}\BibitemShut
  {NoStop}%
\bibitem [{\citenamefont {Hummer}\ and\ \citenamefont
  {{S}zabo}(2001)}]{humm01}%
  \BibitemOpen
  \bibfield  {author} {\bibinfo {author} {\bibfnamefont {G.}~\bibnamefont
  {Hummer}}\ and\ \bibinfo {author} {\bibfnamefont {A.}~\bibnamefont
  {{S}zabo}},\ }\bibfield  {title} {\bibinfo {title} {Free energy
  reconstruction from nonequilibrium single-molecule pulling experiments},\
  }\href@noop {} {\bibfield  {journal} {\bibinfo  {journal} {PNAS}\ }\textbf
  {\bibinfo {volume} {98}},\ \bibinfo {pages} {3658} (\bibinfo {year}
  {2001})}\BibitemShut {NoStop}%
\bibitem [{\citenamefont {Gupta}\ \emph {et~al.}(2011)\citenamefont {Gupta},
  \citenamefont {Abhilash}, \citenamefont {Neupane}, \citenamefont {Yu},
  \citenamefont {Wang},\ and\ \citenamefont {Woodside}}]{gupt11}%
  \BibitemOpen
  \bibfield  {author} {\bibinfo {author} {\bibfnamefont {N.~A.}\ \bibnamefont
  {Gupta}}, \bibinfo {author} {\bibfnamefont {V.}~\bibnamefont {Abhilash}},
  \bibinfo {author} {\bibfnamefont {K.}~\bibnamefont {Neupane}}, \bibinfo
  {author} {\bibfnamefont {H.}~\bibnamefont {Yu}}, \bibinfo {author}
  {\bibfnamefont {F.}~\bibnamefont {Wang}},\ and\ \bibinfo {author}
  {\bibfnamefont {M.~T.}\ \bibnamefont {Woodside}},\ }\bibfield  {title}
  {\bibinfo {title} {Experimental validation of free-energy-landscape
  reconstruction from non-equilibrium single-molecule force spectroscopy
  measurements},\ }\href {https://doi.org/10.1038/NPHYS2022} {\bibfield
  {journal} {\bibinfo  {journal} {Nature Physics}\ }\textbf {\bibinfo {volume}
  {7}},\ \bibinfo {pages} {631} (\bibinfo {year} {2011})}\BibitemShut {NoStop}%
\bibitem [{\citenamefont {Woodside}\ and\ \citenamefont
  {Block}(2014)}]{wood14}%
  \BibitemOpen
  \bibfield  {author} {\bibinfo {author} {\bibfnamefont {M.~T.}\ \bibnamefont
  {Woodside}}\ and\ \bibinfo {author} {\bibfnamefont {S.~M.}\ \bibnamefont
  {Block}},\ }\bibfield  {title} {\bibinfo {title} {{Reconstructing folding
  energy landscapes by single-molecule force spectroscopy}},\ }\href
  {https://doi.org/10.1146/annurev-biophys-051013-022754} {\bibfield  {journal}
  {\bibinfo  {journal} {Annual Review of Biophysics}\ }\textbf {\bibinfo
  {volume} {43}},\ \bibinfo {pages} {19} (\bibinfo {year} {2014})}\BibitemShut
  {NoStop}%
\bibitem [{\citenamefont {Camunas-Soler}\ \emph {et~al.}(2016)\citenamefont
  {Camunas-Soler}, \citenamefont {Ribezzi-Crivellari},\ and\ \citenamefont
  {Ritort}}]{camu16}%
  \BibitemOpen
  \bibfield  {author} {\bibinfo {author} {\bibfnamefont {J.}~\bibnamefont
  {Camunas-Soler}}, \bibinfo {author} {\bibfnamefont {M.}~\bibnamefont
  {Ribezzi-Crivellari}},\ and\ \bibinfo {author} {\bibfnamefont
  {F.}~\bibnamefont {Ritort}},\ }\bibfield  {title} {\bibinfo {title} {{Elastic
  Properties of Nucleic Acids by Single-Molecule Force Spectroscopy}},\ }\href
  {https://doi.org/10.1146/annurev-biophys-062215-011158} {\bibfield  {journal}
  {\bibinfo  {journal} {Annual Review of Biophysics}\ }\textbf {\bibinfo
  {volume} {45}},\ \bibinfo {pages} {65} (\bibinfo {year} {2016})}\BibitemShut
  {NoStop}%
\bibitem [{\citenamefont {Polimeno}\ \emph {et~al.}(2018)\citenamefont
  {Polimeno}, \citenamefont {Magazzù}, \citenamefont {Iatì}, \citenamefont
  {Patti}, \citenamefont {Saija}, \citenamefont {{Esposti Boschi}},
  \citenamefont {Donato}, \citenamefont {Gucciardi}, \citenamefont {Jones},
  \citenamefont {Volpe},\ and\ \citenamefont {Maragò}}]{poli18}%
  \BibitemOpen
  \bibfield  {author} {\bibinfo {author} {\bibfnamefont {P.}~\bibnamefont
  {Polimeno}}, \bibinfo {author} {\bibfnamefont {A.}~\bibnamefont {Magazzù}},
  \bibinfo {author} {\bibfnamefont {M.~A.}\ \bibnamefont {Iatì}}, \bibinfo
  {author} {\bibfnamefont {F.}~\bibnamefont {Patti}}, \bibinfo {author}
  {\bibfnamefont {R.}~\bibnamefont {Saija}}, \bibinfo {author} {\bibfnamefont
  {C.~D.}\ \bibnamefont {{Esposti Boschi}}}, \bibinfo {author} {\bibfnamefont
  {M.~G.}\ \bibnamefont {Donato}}, \bibinfo {author} {\bibfnamefont {P.~G.}\
  \bibnamefont {Gucciardi}}, \bibinfo {author} {\bibfnamefont {P.~H.}\
  \bibnamefont {Jones}}, \bibinfo {author} {\bibfnamefont {G.}~\bibnamefont
  {Volpe}},\ and\ \bibinfo {author} {\bibfnamefont {O.~M.}\ \bibnamefont
  {Maragò}},\ }\bibfield  {title} {\bibinfo {title} {Optical tweezers and
  their applications},\ }\href
  {https://doi.org/https://doi.org/10.1016/j.jqsrt.2018.07.013} {\bibfield
  {journal} {\bibinfo  {journal} {Journal of Quantitative Spectroscopy and
  Radiative Transfer}\ }\textbf {\bibinfo {volume} {218}},\ \bibinfo {pages}
  {131} (\bibinfo {year} {2018})}\BibitemShut {NoStop}%
\bibitem [{\citenamefont {Bustamante}\ \emph {et~al.}(2021)\citenamefont
  {Bustamante}, \citenamefont {Chemla}, \citenamefont {Liu},\ and\
  \citenamefont {Wang}}]{bust21}%
  \BibitemOpen
  \bibfield  {author} {\bibinfo {author} {\bibfnamefont {C.~J.}\ \bibnamefont
  {Bustamante}}, \bibinfo {author} {\bibfnamefont {Y.~R.}\ \bibnamefont
  {Chemla}}, \bibinfo {author} {\bibfnamefont {S.}~\bibnamefont {Liu}},\ and\
  \bibinfo {author} {\bibfnamefont {M.~D.}\ \bibnamefont {Wang}},\ }\bibfield
  {title} {\bibinfo {title} {Optical tweezers in single-molecule biophysics},\
  }\bibfield  {journal} {\bibinfo  {journal} {Nature Reviews Methods Primers}\
  }\textbf {\bibinfo {volume} {1}},\ \href
  {https://doi.org/10.1038/s43586-021-00021-6} {10.1038/s43586-021-00021-6}
  (\bibinfo {year} {2021})\BibitemShut {NoStop}%
\bibitem [{\citenamefont {Gieseler}\ \emph {et~al.}(2021)\citenamefont
  {Gieseler}, \citenamefont {Gomez-Solano}, \citenamefont {Magazz\`{u}},
  \citenamefont {Castillo}, \citenamefont {Garc\'{i}a}, \citenamefont
  {Gironella-Torrent}, \citenamefont {Viader-Godoy}, \citenamefont {Ritort},
  \citenamefont {Pesce}, \citenamefont {Arzola}, \citenamefont
  {Volke-Sep\'{u}lveda},\ and\ \citenamefont {Volpe}}]{gies21}%
  \BibitemOpen
  \bibfield  {author} {\bibinfo {author} {\bibfnamefont {J.}~\bibnamefont
  {Gieseler}}, \bibinfo {author} {\bibfnamefont {J.~R.}\ \bibnamefont
  {Gomez-Solano}}, \bibinfo {author} {\bibfnamefont {A.}~\bibnamefont
  {Magazz\`{u}}}, \bibinfo {author} {\bibfnamefont {I.~P.}\ \bibnamefont
  {Castillo}}, \bibinfo {author} {\bibfnamefont {L.~P.}\ \bibnamefont
  {Garc\'{i}a}}, \bibinfo {author} {\bibfnamefont {M.}~\bibnamefont
  {Gironella-Torrent}}, \bibinfo {author} {\bibfnamefont {X.}~\bibnamefont
  {Viader-Godoy}}, \bibinfo {author} {\bibfnamefont {F.}~\bibnamefont
  {Ritort}}, \bibinfo {author} {\bibfnamefont {G.}~\bibnamefont {Pesce}},
  \bibinfo {author} {\bibfnamefont {A.~V.}\ \bibnamefont {Arzola}}, \bibinfo
  {author} {\bibfnamefont {K.}~\bibnamefont {Volke-Sep\'{u}lveda}},\ and\
  \bibinfo {author} {\bibfnamefont {G.}~\bibnamefont {Volpe}},\ }\bibfield
  {title} {\bibinfo {title} {Optical tweezers --- from calibration to
  applications: a tutorial},\ }\href {https://doi.org/10.1364/AOP.394888}
  {\bibfield  {journal} {\bibinfo  {journal} {Adv. Opt. Photon.}\ }\textbf
  {\bibinfo {volume} {13}},\ \bibinfo {pages} {74} (\bibinfo {year}
  {2021})}\BibitemShut {NoStop}%
\bibitem [{\citenamefont {Stigler}\ \emph {et~al.}(2011)\citenamefont
  {Stigler}, \citenamefont {Ziegler}, \citenamefont {Gieseke}, \citenamefont
  {Gebhardt},\ and\ \citenamefont {Rief}}]{stig11}%
  \BibitemOpen
  \bibfield  {author} {\bibinfo {author} {\bibfnamefont {J.}~\bibnamefont
  {Stigler}}, \bibinfo {author} {\bibfnamefont {F.}~\bibnamefont {Ziegler}},
  \bibinfo {author} {\bibfnamefont {A.}~\bibnamefont {Gieseke}}, \bibinfo
  {author} {\bibfnamefont {J.~C.~M.}\ \bibnamefont {Gebhardt}},\ and\ \bibinfo
  {author} {\bibfnamefont {M.}~\bibnamefont {Rief}},\ }\bibfield  {title}
  {\bibinfo {title} {The complex folding network of single calmodulin
  molecules},\ }\href {https://doi.org/10.1126/science.1207598} {\bibfield
  {journal} {\bibinfo  {journal} {Science}\ }\textbf {\bibinfo {volume}
  {334}},\ \bibinfo {pages} {512} (\bibinfo {year} {2011})}\BibitemShut
  {NoStop}%
\bibitem [{\citenamefont {Satija}\ \emph {et~al.}(2020)\citenamefont {Satija},
  \citenamefont {Berezhkovskii},\ and\ \citenamefont {Makarov}}]{sati20}%
  \BibitemOpen
  \bibfield  {author} {\bibinfo {author} {\bibfnamefont {R.}~\bibnamefont
  {Satija}}, \bibinfo {author} {\bibfnamefont {A.~M.}\ \bibnamefont
  {Berezhkovskii}},\ and\ \bibinfo {author} {\bibfnamefont {D.~E.}\
  \bibnamefont {Makarov}},\ }\bibfield  {title} {\bibinfo {title} {Broad
  distributions of transition-path times are fingerprints of
  multidimensionality of the underlying free energy landscapes},\ }\href
  {https://doi.org/10.1073/pnas.2008307117} {\bibfield  {journal} {\bibinfo
  {journal} {Proceedings of the National Academy of Sciences}\ }\textbf
  {\bibinfo {volume} {117}},\ \bibinfo {pages} {27116} (\bibinfo {year}
  {2020})}\BibitemShut {NoStop}%
\bibitem [{\citenamefont {Berezhkovskii}\ and\ \citenamefont
  {Makarov}(2018)}]{bere18}%
  \BibitemOpen
  \bibfield  {author} {\bibinfo {author} {\bibfnamefont {A.~M.}\ \bibnamefont
  {Berezhkovskii}}\ and\ \bibinfo {author} {\bibfnamefont {D.~E.}\ \bibnamefont
  {Makarov}},\ }\bibfield  {title} {\bibinfo {title} {Single-molecule test for
  markovianity of the dynamics along a reaction coordinate},\ }\href
  {https://doi.org/10.1021/acs.jpclett.8b00956} {\bibfield  {journal} {\bibinfo
   {journal} {The Journal of Physical Chemistry Letters}\ }\textbf {\bibinfo
  {volume} {9}},\ \bibinfo {pages} {2190} (\bibinfo {year} {2018})}\BibitemShut
  {NoStop}%
\bibitem [{\citenamefont {Lapolla}\ and\ \citenamefont {Godec}(2021)}]{lapo21}%
  \BibitemOpen
  \bibfield  {author} {\bibinfo {author} {\bibfnamefont {A.}~\bibnamefont
  {Lapolla}}\ and\ \bibinfo {author} {\bibfnamefont {A.~c.~v.}\ \bibnamefont
  {Godec}},\ }\bibfield  {title} {\bibinfo {title} {Toolbox for quantifying
  memory in dynamics along reaction coordinates},\ }\href
  {https://doi.org/10.1103/PhysRevResearch.3.L022018} {\bibfield  {journal}
  {\bibinfo  {journal} {Phys. Rev. Res.}\ }\textbf {\bibinfo {volume} {3}},\
  \bibinfo {pages} {L022018} (\bibinfo {year} {2021})}\BibitemShut {NoStop}%
\bibitem [{\citenamefont {Spinney}\ and\ \citenamefont {Ford}(2012)}]{spin12}%
  \BibitemOpen
  \bibfield  {author} {\bibinfo {author} {\bibfnamefont {R.~E.}\ \bibnamefont
  {Spinney}}\ and\ \bibinfo {author} {\bibfnamefont {I.~J.}\ \bibnamefont
  {Ford}},\ }\bibfield  {title} {\bibinfo {title} {Nonequilibrium
  thermodynamics of stochastic systems with odd and even variables},\ }\href
  {https://doi.org/10.1103/PhysRevLett.108.170603} {\bibfield  {journal}
  {\bibinfo  {journal} {PRL}\ }\textbf {\bibinfo {volume} {108}},\ \bibinfo
  {pages} {170603} (\bibinfo {year} {2012})}\BibitemShut {NoStop}%
\bibitem [{\citenamefont {Dieball}\ and\ \citenamefont {Godec}(2022)}]{dieb22}%
  \BibitemOpen
  \bibfield  {author} {\bibinfo {author} {\bibfnamefont {C.}~\bibnamefont
  {Dieball}}\ and\ \bibinfo {author} {\bibfnamefont {A.}~\bibnamefont
  {Godec}},\ }\bibfield  {title} {\bibinfo {title} {Mathematical,
  thermodynamical, and experimental necessity for coarse graining empirical
  densities and currents in continuous space},\ }\href
  {https://doi.org/10.1103/PhysRevLett.129.140601} {\bibfield  {journal}
  {\bibinfo  {journal} {Phys. Rev. Lett.}\ }\textbf {\bibinfo {volume} {129}},\
  \bibinfo {pages} {140601} (\bibinfo {year} {2022})}\BibitemShut {NoStop}%
\end{thebibliography}%

\end{document}